\documentclass[fleqn,10pt]{wlscirep}
\usepackage[utf8]{inputenc}
\usepackage[T1]{fontenc}

\usepackage{subcaption}
\usepackage{enumitem}
\usepackage{multirow}

\title{Exposure to Content Written by Large Language Models Can Reduce Stigma Around Opioid Use Disorder in Online Communities}

\author[1,*]{Shravika Mittal}
\author[1]{Darshi Shah}
\author[1]{Shin Won Do}
\author[2]{Mai ElSherief}
\author[3]{Tanushree Mitra}
\author[1]{Munmun De Choudhury}
\affil[1]{College of Computing, Georgia Institute of Technology, Atlanta, Georgia, USA}
\affil[2]{Khoury College of Computer Science, Northeastern University, Boston, Massachusetts, USA}
\affil[3]{Information School, University of Washington, Seattle, Washington, USA}
\affil[*]{smittal87@gatech.edu}

\begin{abstract}

Widespread stigma, both in the offline and online spaces, acts as a barrier to harm reduction efforts in the context of opioid use disorder (OUD). This stigma is prominently directed towards clinically approved medications for addiction treatment (MAT), people with the condition, and the condition itself. Given the potential of artificial intelligence based technologies in promoting health equity, and facilitating empathic conversations, this work examines whether large language models (LLMs) can help abate OUD-related stigma in online communities. To answer this, we conducted a series of pre-registered randomized controlled experiments, where participants read LLM-generated, human-written, or no responses to help seeking OUD-related content in online communities. The experiment was conducted under two setups, i.e., participants read the responses either once (N = $2,141$), or repeatedly for 14 days (N = $107$). We found that participants reported the least stigmatized attitudes toward MAT after consuming LLM-generated responses under both the setups. This study offers insights into strategies that can foster inclusive online discourse on OUD, e.g., based on our findings LLMs can be used as an education-based intervention to promote positive attitudes and increase people's propensity toward MAT.

\end{abstract}
\begin{document}

\flushbottom
\maketitle
%
%
\thispagestyle{empty}


\section*{Introduction}

In the U.S., opioid overdose continues to be a leading cause of death. The National Center for Health Statistics at the U.S. Centers for Disease Control and Prevention reported that opioid-involved overdose deaths rose from $49,860$ in 2019 to $81,806$ in 2022~\cite{nihDrugOverdose}. Addressing this crisis requires reducing barriers to treatment, which manifest as widespread stigma~\cite{link2001conceptualizing} in the offline settings targeting the condition~\cite{doi:10.1080/10826084.2022.2079133,richard2020you,ADAMS2021321}, i.e., opioid use disorder (OUD), people with OUD~\cite{barry2014stigma,Taylor15102021,corrigan2009public}, and clinically approved medications for addiction treatment (MAT)~\cite{doi:10.1177/1178221816685087,MADDEN2019324,richard2020you}. To avoid such social, structural, and intervention-based marginalization, people with OUD often seek non-conventional, online pathways to recovery. Owing to pseudonymity, online communities, such as those on Reddit, are particularly popular among people with OUD~\cite{balsamo2023pursuit,chancellor2019discovering}. Individuals use these forums to freely discuss opioid use~\cite{almeida2024use,park2018opioid}, to explore alternative treatment frameworks~\cite{chancellor2019discovering}, and to offer and seek support around issues relating to recovery~\cite{balsamo2023pursuit,bunting2021socially}. 

However, online communities are not necessarily always the safe spaces they aspire to be. They are home to vast amounts of unhelpful advice, misleading, and clinically unverified content,~\cite{chen2022infodemic,sager2021identifying,pollack2022characterizing,suarez2021prevalence} and can exacerbate the marginalization of already vulnerable populations~\cite{thach2024visible,10.1145/3686940,10.1145/3512901}. Platforms such as Reddit are known to contain stigmatized attitudes toward MAT~\cite{elsherief2021characterizing,doi:10.1080/10826084.2022.2091786}  and can reinforce harmful stereotypes by normalizing language that labels people with OUD as `addicts' or frames their condition as a personal failing~\cite{Giorgi_Bellew_Habib_Sedoc_Smitterberg_Devoto_Himelein-Wachowiak_Curtis_2024}. It is imperative to improve this online ecosystem by fostering an environment that reduces stigma, as for people with OUD, such platforms may be their primary or sole source of information, support, and validation.

With advancements in artificial intelligence (AI) based technologies, scholars have started to investigate the potential of AI to augment human capabilities across a wide range of creative, complex, and high-risk tasks -- e.g., using text-based large language models (LLMs) to promote health equity~\cite{pierson2025using}, to facilitate empathic conversations in peer to peer mental health support~\cite{sharma2023human,10.1145/3654777.3676430,hsu2023helping}, to provide mental health treatment~\cite{doi:10.1056/AIoa2400802}, and to motivate behavior change~\cite{jorke2024supporting}. Inspired by these, this paper examines the potential of LLMs in writing responses to help seeking OUD content in online communities. Broadly, we seek to understand if LLMs can support ongoing efforts toward mitigating OUD-related stigma 
in these online spaces. 

To do so, we conducted a series of pre-registered 
randomized controlled experiments. 
The three most prevalent sources known to amplify OUD-related stigma in online communities are negative attitudes toward
(a) MAT~\cite{elsherief2021characterizing,icwsm2025,doi:10.1080/10826084.2022.2091786}, which is framed as merely replacing one drug with another; (b) people with OUD~\cite{icwsm2025}, who are portrayed as `dangerous' or `weak'; and (c) OUD itself~\cite{info:doi/10.2196/44726}, with full recovery from the condition being represented as impossible. Therefore, we operationalized these three attitudes as our dependent variables (DVs). In our experiments, we assessed whether LLM-written content could reduce stigmatized attitudes toward MAT (DV1), people with OUD (DV2), and OUD (DV3). 
Using a between-subjects study design, participants were randomly assigned to one of three interventions: (a) LLM, participants read LLM-generated responses to online queries on OUD (sourced from Reddit); (b) Human, participants read human-written responses (again, sourced from Reddit) to the same set of queries as the LLM intervention group; and (c) Control, participants were not provided any content to read.
We hypothesized that the LLM intervention would reduce stigmatized attitudes, across all three DVs, more effectively than the no intervention (i.e., Control) and Human intervention conditions. 

The experiment was conducted under two setups: (a) single exposure (N = $2,141$); where participants interacted with the intervention, i.e., read the relevant responses, once and (b) longitudinal exposure (N = $107$); where participants interacted with the intervention daily for 14 consecutive days, with the content varying each day. The single exposure setup was inspired by Schleider et al.'s~\cite{info:doi/10.2196/20513} conceptualization of single-session interventions, which are structured programs involving only one encounter with a clinic, provider, or program. Prior work supports utility of such short-term interventions for promoting behavior change, particularly in the context of anxiety, depression, and substance use~\cite{schleider2025single}. Moreover, the behavioral psychology literature suggests that repeated, longitudinal exposure can further enhance attitude change~\cite{annurev:/content/journals/10.1146/annurev-psych-122216-011911}. By evaluating interventions under these two setups, the paper seeks to gain insights into the differential impacts and capabilities of brief, one-time and prolonged, repeated interventions, providing valuable guidance for real-world deployments. 

Our hypothesis was strongly supported for DV1, under both single and longitudinal exposure setups. Specifically, the LLM intervention was the most effective at reducing stigmatized attitudes toward MAT compared to the Human and Control interventions. That said, certain interventions had a backfire effect~\cite{swire2022backfire}. A single exposure to human-written responses led to worsened attitudes across all three DVs, even when compared to the no intervention Control. Lastly, the interventions did not affect all participants equally. They had a detrimental impact on participants with highly approving pre-intervention attitudes but were effective in reducing stigma among those with more stigmatizing pre-intervention attitudes.
\section*{Results}

\subsection*{Findings of Hypotheses Testing}

\subsubsection*{H1(a): LLM intervention would reduce stigmatized attitudes toward MAT (DV1) to a greater extent compared to the no intervention Control.}

Our study revealed that H1(a) was supported for both the single and longitudinal exposure setups, with statistical significance (Tables~\ref{tab:model_single-shot} and~\ref{tab:model_repeated-exp}). Compared to the no intervention condition, i.e., Control, participants in the LLM intervention reported more approving change in attitudes, or reduced stigma, toward MAT (Figure~\ref{fig:res_1}), for both the single ($0.248$ vs. $0.113$, $p$: $2.64 \times 10^{-9}$) and longitudinal ($0.435$ vs. $0.092$, $p$: $0.00269$) exposure setups. 

\subsubsection*{H1(b): LLM intervention would reduce stigmatized attitudes toward MAT (DV1) to a greater extent compared to the Human intervention.} 

H1(b) was supported for both the single and longitudinal exposure setups, with statistical significance (Tables~\ref{tab:model_single-shot} and~\ref{tab:model_repeated-exp}). Compared to the Human intervention, participants in the LLM condition reported a more approving change in attitudes (Figure~\ref{fig:res_1}) toward MAT, for both the single ($0.248$ vs. $0.008$, $p$ < $2.00 \times 10^{-16}$) and longitudinal ($0.435$ vs. $0.048$, $p$ < $0.00136$) exposure setups.

\subsubsection*{H2(a): LLM intervention would reduce stigmatized attitudes toward people with OUD (DV2) to a greater extent compared to the no intervention Control.} 

Compared to the Control, participants in the LLM intervention reported more approving change in attitudes toward people with OUD (Figure~\ref{fig:res_1}, Tables~\ref{tab:model_single-shot} and~\ref{tab:model_repeated-exp}), for both the single ($0.058$ vs. $0.022$, $p$: $0.488$) and longitudinal ($-0.247$ vs. $-0.419$, $p$: $0.131$) exposure setups.

\subsubsection*{H2(b): LLM intervention would reduce stigmatized attitudes toward people with OUD (DV2) to a greater extent compared to the Human intervention.} 

Compared to the Human intervention, participants in the LLM intervention reported more approving change in attitudes toward people with OUD (Figure~\ref{fig:res_1}, Tables~\ref{tab:model_single-shot} and~\ref{tab:model_repeated-exp}) for both the single ($0.058$ vs. $-0.112$, $p$: $3.40 \times 10^{-12}$) and longitudinal ($-0.247$ vs. $-0.303$, $p$: $0.633$) exposure setups, with H2(b) being supported in the single exposure setup.

\subsubsection*{H3(a): LLM intervention would reduce stigmatized attitudes toward OUD (DV3) to a greater extent compared to the no intervention Control.}

Single exposure to the LLM intervention increased stigmatizing attitudes toward OUD when compared to the Control condition ($-0.229$ vs. $-0.164$, $p$: $0.005$; Figure~\ref{fig:res_1_single-shot}). Therefore, H3(a) was rejected, with statistical significance, in the single exposure setup (Table~\ref{tab:model_single-shot}). In the longitudinal exposure setup, LLM intervention led to more improvements in participant attitudes toward OUD, in comparison to the Control ($0.193$ vs. $-0.022$, $p$: $0.055$; Figure~\ref{fig:res_1_repeated-exposure}).

\subsubsection*{H3(b): LLM intervention would reduce stigmatized attitudes toward OUD (DV3) to a greater extent compared to the Human intervention.}

Compared to the Human intervention, participants in the LLM intervention reported more approving change in attitudes toward OUD (Figure~\ref{fig:res_1}, Tables~\ref{tab:model_single-shot} and~\ref{tab:model_repeated-exp}) for both the single ($-0.229$ vs. $-0.437$, $p$ < $2.00 \times 10^{-16}$) and longitudinal ($0.193$ vs. $0.058$, $p$: $0.247$) exposure setups, with H3(b) being supported in the single exposure setup.

\subsubsection*{Varied Impact on Post-Intervention Outcomes by Dependent Variable, Exposure Setup, and Intervention Type}

Among the three DVs, the LLM intervention was the most successful at reducing stigmatizing attitudes toward MAT, i.e., DV1 (Figure~\ref{fig:res_1}). On average, after a single exposure to LLM-generated responses, participants reported an improvement of $6.61$\% in their perceptions toward MAT. This approving change in attitudes was even more pronounced in the longitudinal exposure setup ($13.64$\%). 

While the longitudinal exposure setup outperformed the single exposure setup in improving participant attitudes toward MAT and OUD, as seen post both Human (DV1: $0.008$ (single) vs. $0.048$ (longitudinal); DV3: $-0.437$ vs. $0.058$) and LLM (DV1: $0.248$ vs. $0.435$; DV3: $-0.229$ vs. $0.193$) interventions (Tables~\ref{tab:model_single-shot} and~\ref{tab:model_repeated-exp}), it was not always better. For DV2, the longitudinal exposure actually worsened stigmatizing attitudes toward people with OUD compared to the single exposure setup, post both Human ($-0.112$ vs. $-0.303$) and LLM ($0.058$ vs. $-0.247$) interventions.

Interventions can sometimes have a backfire effect~\cite{swire2022backfire}. In our case, certain interventions amplified stigmatizing attitudes, post intervention, in comparison to the no intervention (i.e., Control) baseline (Figure~\ref{fig:res_1_single-shot}). Specifically, a single exposure to the Human intervention increased stigmatized perceptions by $0.99$\%, $4.73$\%, and $11.58$\% toward MAT, people with OUD, and OUD, respectively, which was worse compared to the change in attitudes after no intervention at all (stigmatized perceptions decreased by $1.82$\% toward MAT, and increased by $0.28$\% and $1.26$\% toward people with OUD and OUD, respectively). Although we observed a change in attitudes for participants in the Control condition, this change was not statistically significant under Kruskal Wallis H-tests (Table~\ref{tab:control-anova}).

\subsection*{Analysis of the Intervention Content}
\begin{figure*}
\centering

    \begin{subfigure}[b]{0.45\textwidth}
    \centering
    \includegraphics[width=\textwidth]{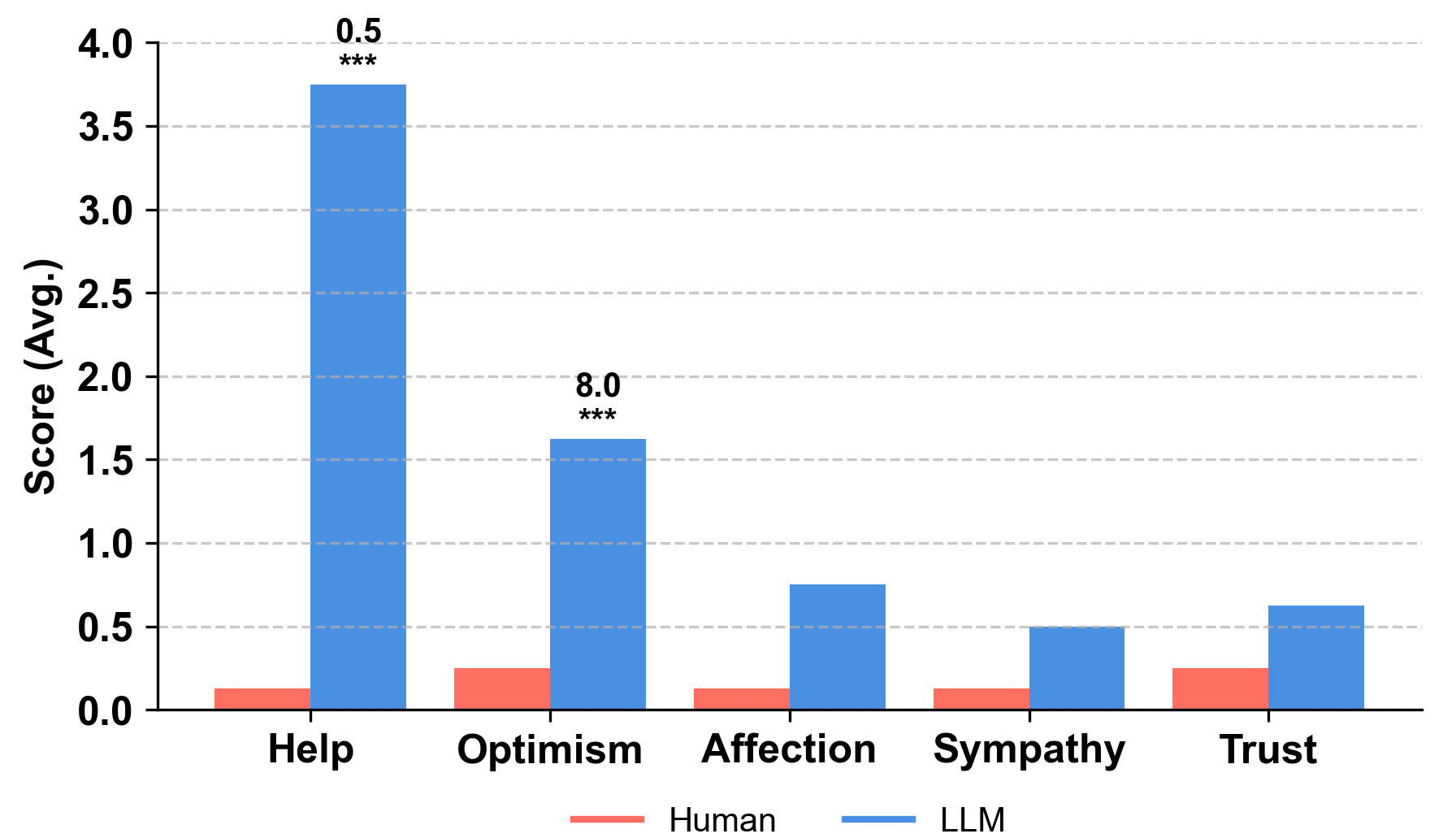}
    \caption{Emotional appeal (Single exposure setup)} \label{fig:single-empath}
    \end{subfigure}
    \hfill
    \begin{subfigure}[b]{0.45\textwidth}
    \centering
    \includegraphics[width=\textwidth]{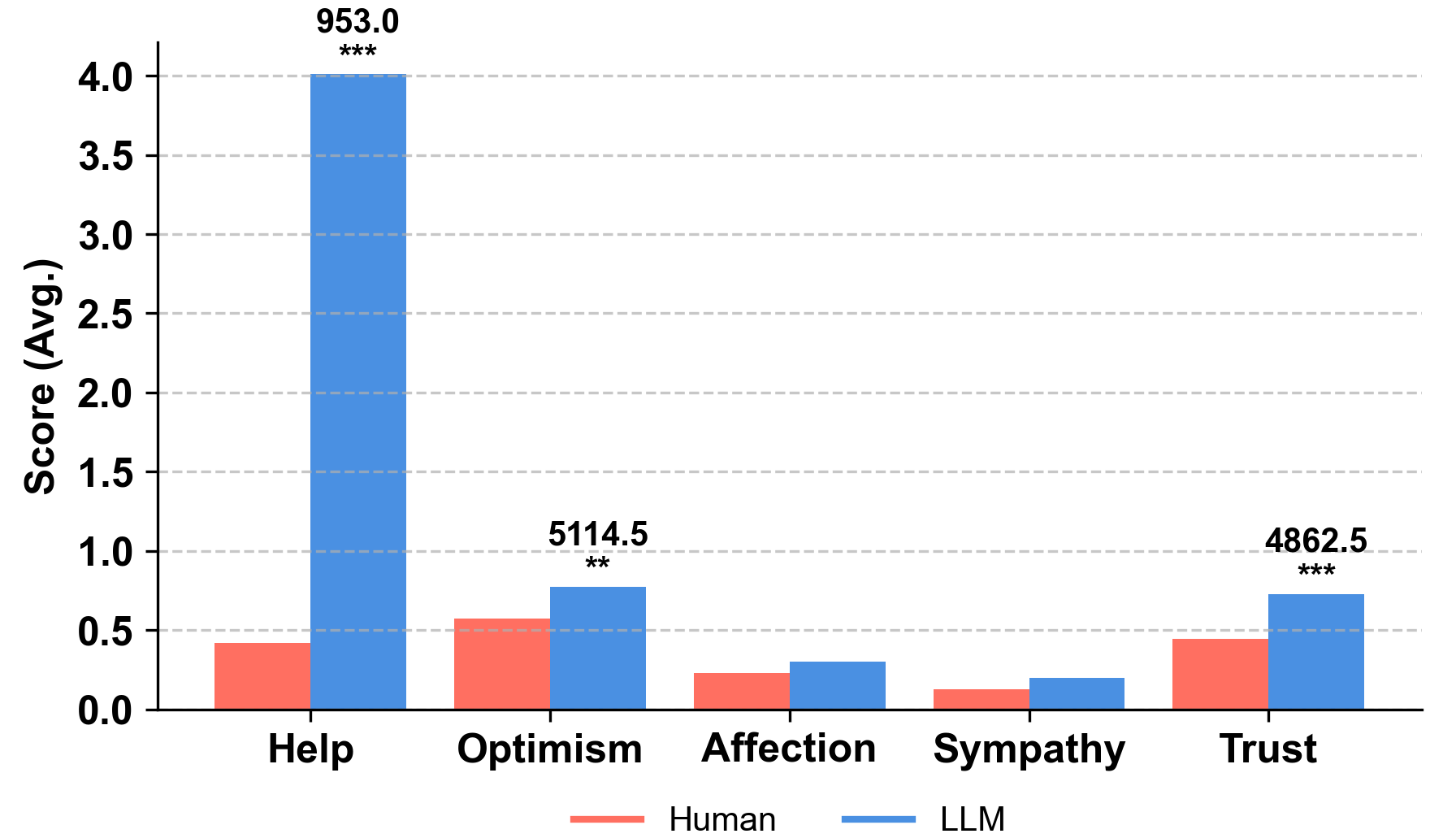}
    \caption{Emotional appeal (Longitudinal exposure setup)} \label{fig:longitudinal-empath}
    \end{subfigure}
    \hfill

    \begin{subfigure}[b]{0.45\textwidth}
    \centering
    \includegraphics[width=\textwidth]{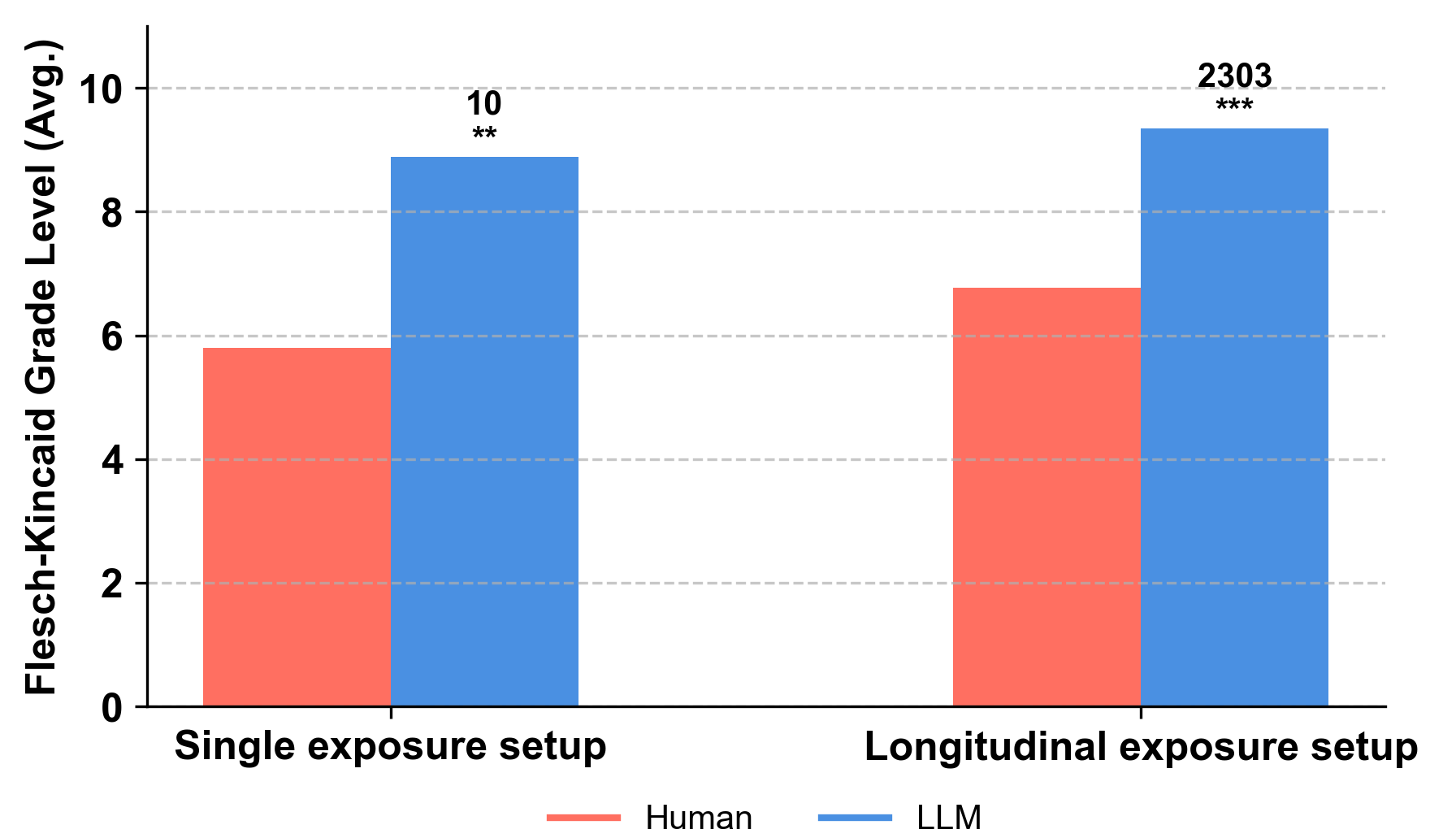}
    \caption{Readability} \label{fig:readability}
    \end{subfigure}
    \hfill
    \begin{subfigure}[b]{0.45\textwidth}
    \centering
    \includegraphics[width=\textwidth]{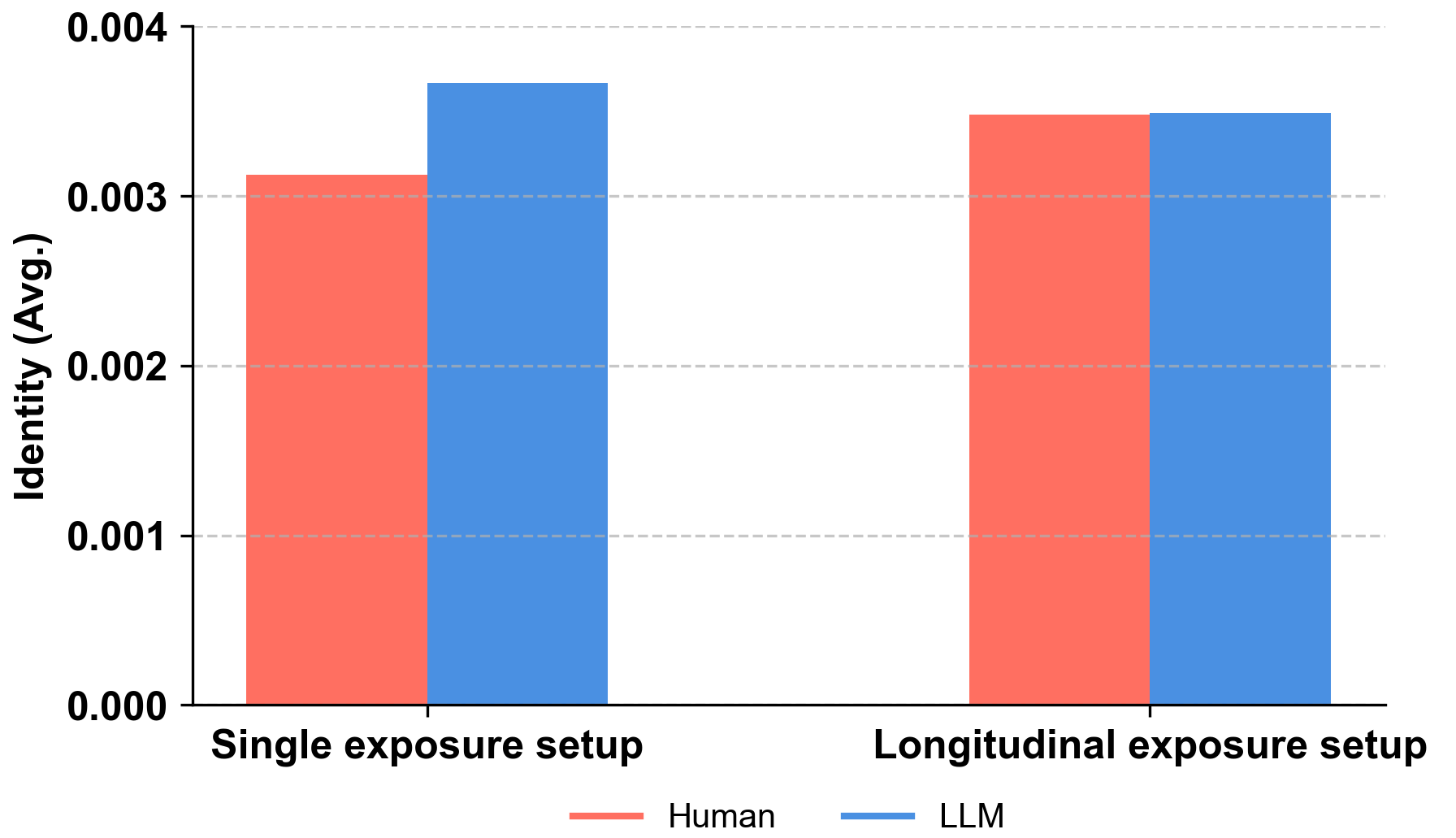}
    \caption{Shared sense of belonging} \label{fig:identity}
    \end{subfigure}
    \hfill

    \begin{subfigure}[b]{0.45\textwidth}
    \centering
    \includegraphics[width=\textwidth]{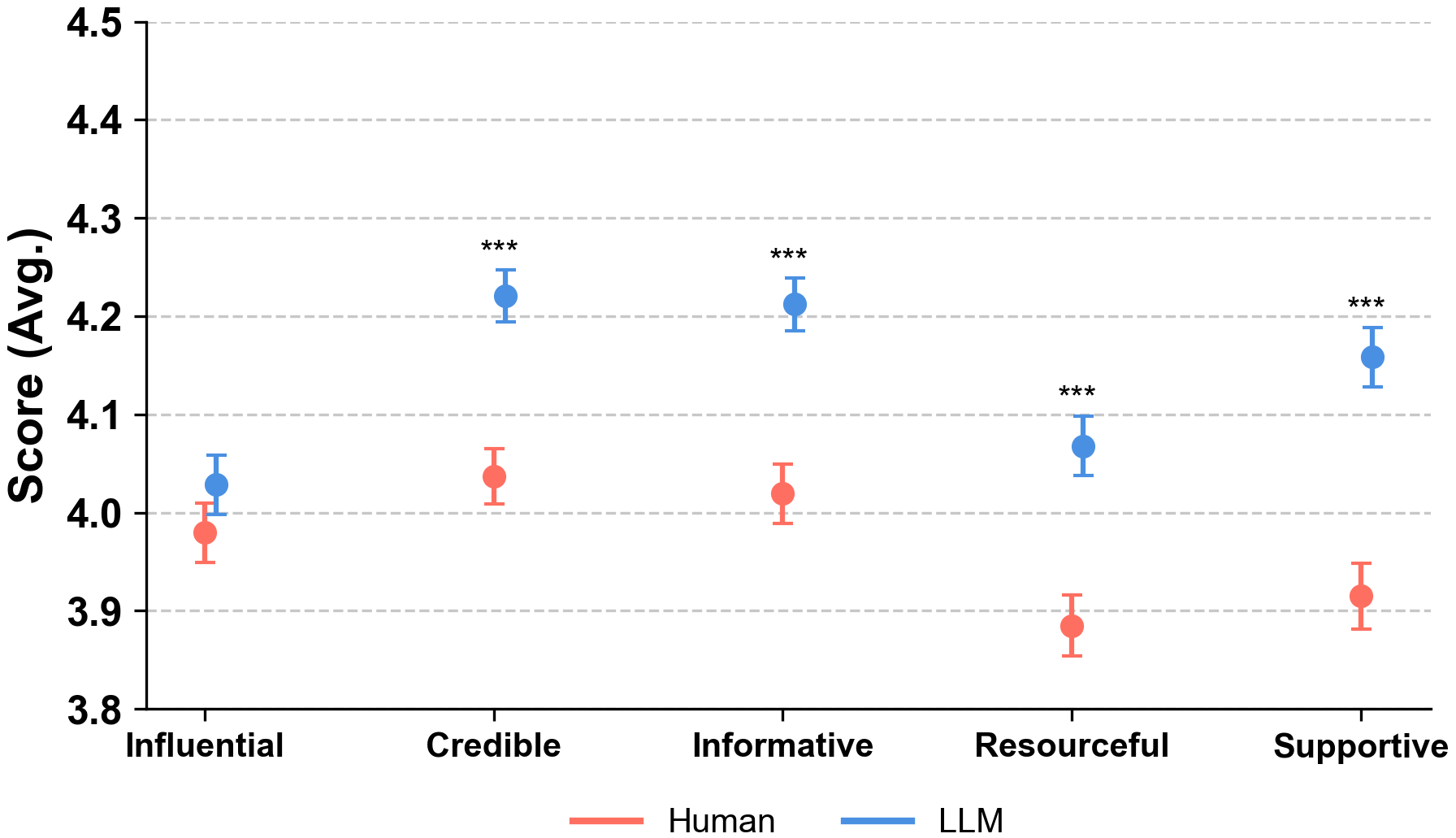}
    \caption{Participant attitudes} \label{fig:DV4}
    \end{subfigure}
    \hfill
    
    \caption{\emph{Analysis of the intervention content.} Responses read by participants within the LLM and Human intervention groups were evaluated for \textbf{(a)}, \textbf{(b)} emotional appeal, \textbf{(c)} readability, and \textbf{(d)} shared sense of belonging. Emotional appeal is reported using five relevant categories available in Empath~\cite{fast2016empath}, a lexicon-based tool; a higher score is indicative of a higher alignment to the category. Readability is reported using the Flesch-Kincaid Grade Level index~\cite{kincaid1975derivation}; a lower score is indicative of simpler, more readable text. Shared sense of belonging is reported using the identity social dimension classifier~\cite{choi2020ten}, which quantifies in-group or community forward linguistic cues; the higher the score the better. Scores are averaged across all the responses read by participants during the single and longitudinal exposure setups. Mann-Whitney U-tests were performed to explore differences in score distributions for responses provided in the LLM and Human interventions. Statistically significant differences are noted with the test statistic and p-values ($p$): * ($p$ < $0.05$), ** ($p$ < $0.01$), or *** ($p$ < $0.001$). \\
    \textbf{(e)} At the end of our experiments, participants in the LLM and Human interventions rated the responses consumed during the respective interventions, using a 5-point Likert scale, as: (1) influential, \textit{responses offered a different approach to look at OUD}; (2) credible, \textit{responses were reasonable and trustworthy}; (3) informative, \textit{responses were knowledgeable}; (4) resourceful, \textit{likely to refer to such responses to gain information about OUD}; and (5) supportive, \textit{prefer to receive such responses if one had OUD}. On finding no significant differences in ratings across single and longitudinal exposure setups, we combined participant ratings for the two setups and report a weighted average (weighted by the sample size). Mann-Whitney U-tests were performed to examine differences in score distributions for ratings provided by participants in the LLM and Human intervention groups. Statistically significant differences are noted with the p-values ($p$): * ($p$ < $0.05$), ** ($p$ < $0.01$), or *** ($p$ < $0.001$).}
\label{fig:content_analysis}
\end{figure*}

In the context of behavioral health conditions, Link and Phelan highlight the potential of simple interventions that elicit hope, optimism, and shared sense of belonging in reducing stigma~\cite{link2001conceptualizing}. Consequently, we examined the intervention content, read by participants in the LLM and Human interventions, across linguistic dimensions such as readability, emotional appeal, and in-group affinity to provide insights into the rationale behind the effectiveness of the various interventions reported above. We used relevant categories from Empath~\cite{fast2016empath}, a lexicon-based tool, to assess emotional appeal (Figures~\ref{fig:single-empath} and~\ref{fig:longitudinal-empath}).  For both the single and longitudinal exposure setups, the LLM intervention content contained significantly more optimistic (`optimism' category) and supportive (`help' category) linguistic cues compared to the Human intervention content, which are indicative of encouraging, hopeful, and positive discourse. Next, though not significantly, the LLM intervention content contained a greater shared sense of belonging (Figure~\ref{fig:identity}), which was quantified using the identity social dimension classifier~\cite{choi2020ten} widely used to determine peer support capabilities of online communities. Again, this provides empirical evidence suggesting that the LLM intervention contained more inclusive linguistic cues, potentially helpful to abate stigma~\cite{link2001conceptualizing}. Further, post intervention, participants in the LLM and Human intervention groups self-reported their attitudes toward the intervention content (Table~\ref{tab:DV4_survey}), i.e., whether they found the responses to be influential; \textit{offered a different approach to look at OUD}, credible; \textit{were reasonable and trustworthy}, informative; \textit{were knowledgeable}, resourceful; \textit{likely to refer to such responses to gain information about OUD}, and supportive; \textit{prefer to receive such responses if one had OUD}. On finding no significant differences in ratings across single and longitudinal exposure setups, we combined participant scores for the two setups using a weighted average measure (weighted by the sample size; Figure~\ref{fig:DV4}). Participants in the LLM intervention rated the LLM-generated responses as more influential, and significantly more credible, informative, resourceful, and supportive compared to how participants in the Human intervention rated the human-written responses. 

\subsection*{Participant Perspectives on Intervention Content}

Next, we explored thematic patterns within participants' text-based post-intervention reflections, which they provided daily during the longitudinal exposure setup. The reflections were participants' responses to open-ended questions such as `Based on your reading, is Methadone the best route to recovery?' and `Is full recovery possible?' We adopted a mixed-methods approach to manually annotate themes within a handful of reflections, which were then used as ground truth to train LLM-based classifiers for machine-annotating the rest (refer to~\ref{sec:thematic_analysis} for details). The two most prominent themes relevant to DV1 in participants' reflections were `Dependence', i.e., MAT replaces one drug with another, and `Effectiveness', i.e., MAT is an effective treatment for OUD. Recurring themes related to DV2 were `Blame', meaning people with OUD are responsible for their own condition and `Labeling', referring to the use of derogatory terms like `addict' and `junkies'. For DV3, the two most prominent themes were `Fatality' (i.e., full recovery from OUD is not possible) and its contrast, `Belief in Recovery'. Refer to Tables~\ref{tab:codebook_DV1},~\ref{tab:codebook_DV2}, and~\ref{tab:codebook_DV3} for a complete list of themes. Per DV1, more reflections from participants in the LLM intervention ($n = 53/102$) supported the effectiveness of MAT compared to those from participants in the Human intervention ($n = 46/102$). Fewer reflections from participants in the LLM intervention highlighted dependence on MAT ($n = 38/102$), and such reflections were also less in comparison to those from participants in the Human condition ($n = 39/102$). Though the two most recurring themes for DV2 highlight negative attitudes toward people with OUD, the support for them was less pronounced among participants' reflections in the LLM intervention compared to those in the Human intervention (`Blame': $n = 30/136$ (LLM) vs. $n = 31/136$ (Human); `Labeling': $37/136$ vs. $50/136$). In the LLM intervention, many more reflections emphasized the belief in recovery ($n = 80/102$) compared to those rejecting it ($n = 35/102$). Again, this positive support was more evident in reflections from participants in the LLM intervention compared to those in the Human intervention. Only $69/102$ reflections from participants in the Human intervention supported the belief in recovery while $46/102$ dismissed it.

\subsection*{Post-hoc Analyses}

\subsubsection*{Impact of Pre-intervention Attitudes on Post-intervention Outcomes} 
\begin{figure*}
\centering
    
    \par\vspace{1em}  
    \textbf{Single exposure setup}

    \begin{subfigure}[b]{0.33\textwidth}
    \centering
    \includegraphics[width=\textwidth]{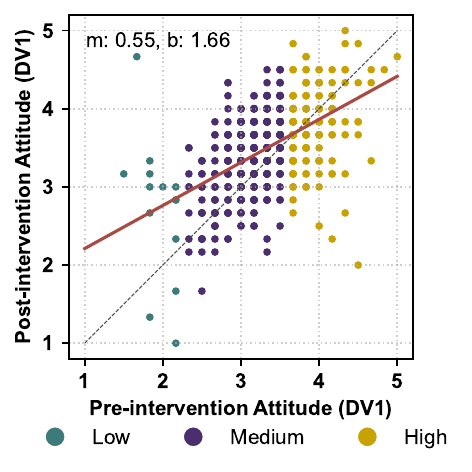}
    \caption{} \label{fig:single-shot_scatter_llm_DV1}
    \end{subfigure}
    \hfill
    \begin{subfigure}[b]{0.33\textwidth}
    \centering
    \includegraphics[width=\textwidth]{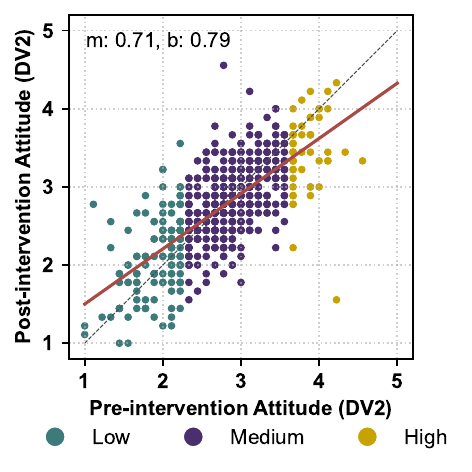}
    \caption{} \label{fig:single-shot_scatter_llm_DV2}
    \end{subfigure}
    \hfill
    \begin{subfigure}[b]{0.33\textwidth}
    \centering
    \includegraphics[width=\textwidth]{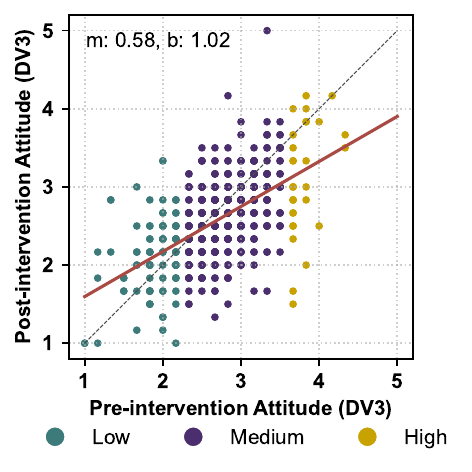}
    \caption{} \label{fig:single-shot_scatter_llm_DV3}
    \end{subfigure}
    \hfill

    \par\vspace{1em}  
    \textbf{Longitudinal exposure setup}
    
    \begin{subfigure}[b]{0.33\textwidth}
    \centering
    \includegraphics[width=\textwidth]{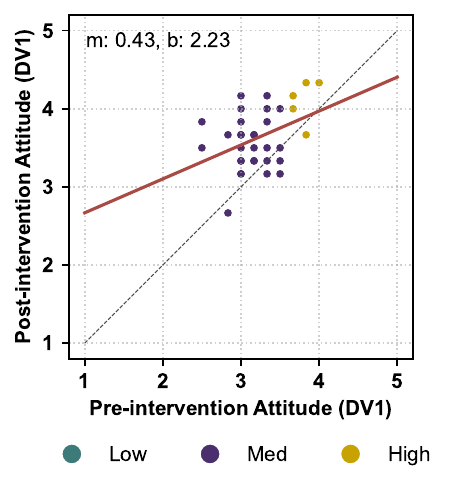}
    \caption{} \label{fig:repeated-exp_scatter_llm_DV1}
    \end{subfigure}
    \begin{subfigure}[b]{0.33\textwidth}
    \centering
    \includegraphics[width=\textwidth]{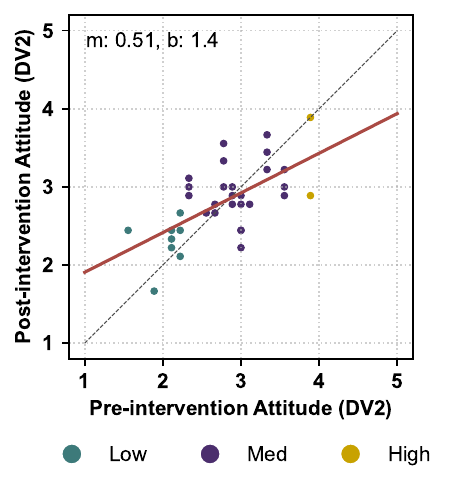}
    \caption{} \label{fig:repeated-exp_scatter_llm_DV2}
    \end{subfigure}
    \begin{subfigure}[b]{0.33\textwidth}
    \centering
    \includegraphics[width=\textwidth]{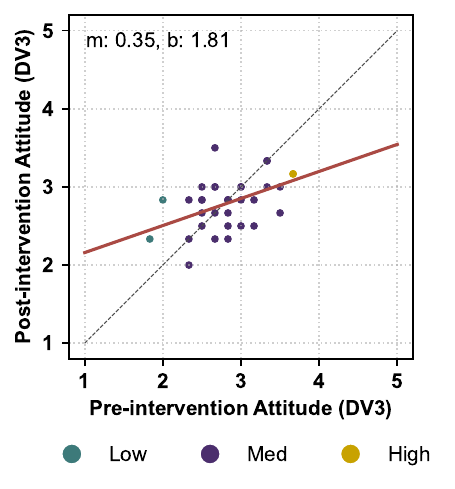}
    \caption{} \label{fig:repeated-exp_scatter_llm_DV3}
    \end{subfigure}
    \caption{\emph{Pre- and post-intervention attitudes toward the three DVs for participants within the LLM intervention.} (a), (b), and (c): single exposure setup; (d), (e), and (f): longitudinal exposure setup. Participants were divided into three groups -- low, medium, and high -- based on their pre-intervention attitudes. The Likert scale range (1 to 5) was equally divided into three parts to achieve this. The red lines indicate the regression line or the best linear fit for the pre-/post-intervention score distributions (m: slope, b: intercept). The gray dotted line represents the no-change post-intervention fit, i.e., $Y_{pre} = Y_{post}$. Participants (represented via dots) above (below) the gray line reported a higher (lower) attitude post intervention. Participants in the low (high) pre-intervention score category were more concentrated in the region above (below) the no-change post-intervention line fit.}
\label{fig:scatter-plots-llm}
\end{figure*}

For participants within the LLM condition, we further examined whether the intervention impacted them all similarly (Figure~\ref{fig:scatter-plots-llm}). To do so, we divided the participants into 3 groups, low: [1, 2.33), medium: [2.33, 3.66), and high: [3.66, 5], based on their pre-intervention attitudes toward each DV. Participants in the low pre-intervention attitude category were more likely to report a boost in their post-intervention attitudes. This was consistent across both single (DV1: $N = 14$, paired pre-/post- T-test statistic ($t$-stat): $-2.73$, $p$: $0.017$; DV2: $N = 133$, $t$-stat: $-4.64$, $p$: $8.27 \times 10^{-6}$; DV3: $N = 100$, $t$-stat: $-4.19$, $p$: $6.17 \times 10^{-5}$) and longitudinal (DV2: $N = 8$, $t$-stat: $-1.94$, $p$: $0.09$; DV3: $N = 2$, $t$-stat: $-4.00$, $p$: $0.16$) exposure setups. Participants in the high pre-intervention score category reported the opposite; they were more likely to exhibit a decrease in their attitudes post intervention -- again, for both single (DV1: $N = 171$, $t$-stat: $3.55$, $p$: $0.0005$; DV2: $N = 51$, $t$-stat: $5.63$, $p$: $8.27 \times 10^{-7}$; DV3: $N = 31$, $t$-stat: $5.98$, $p$: $1.46 \times 10^{-6}$) and longitudinal (DV1: $N = 5$, $t$-stat: $-2.45$, $p$: $0.07$; DV2: $N = 2$, $t$-stat: $1.00$, $p$: $0.49$) exposure setups, with one exception. This suggests that participants with stigmatized pre-intervention attitudes are more likely to adopt approving perspectives following the intervention. Contrastively, the intervention can be detrimental for participants with already highly approving pre-intervention attitudes. This remained unchanged when we accounted for demographics such as age, gender, education level, and political leaning (Table~\ref{tab:sm_t-test_demographics_llm}). The finding also holds true for participants enrolled in the Human intervention (Figure~\ref{fig:scatter-plots-human}, Table~\ref{tab:sm_t-test_demographics_human}).

\subsubsection*{Impact of External Information Consumption on Post-intervention Outcomes}

We performed a secondary analysis to examine whether participants' external consumption of OUD-related information impacted their post-intervention attitudes. To assess this, we added an additional fixed effect to our original linear mixed-effects model (\ref{sec:lmm}), which captured participants' self-reported response to whether or not they consumed information on OUD beyond our provided interventions (\ref{sec:impact_ext_engagement}). This external consumption did not have a statistically significant impact on post-intervention change in attitudes across all three DVs and the two exposure setups (Table~\ref{tab:ext_engagement}).
\section*{Discussion}

\begin{figure}
    \begin{subfigure}{0.45\textwidth}
        \includegraphics[width=\linewidth]{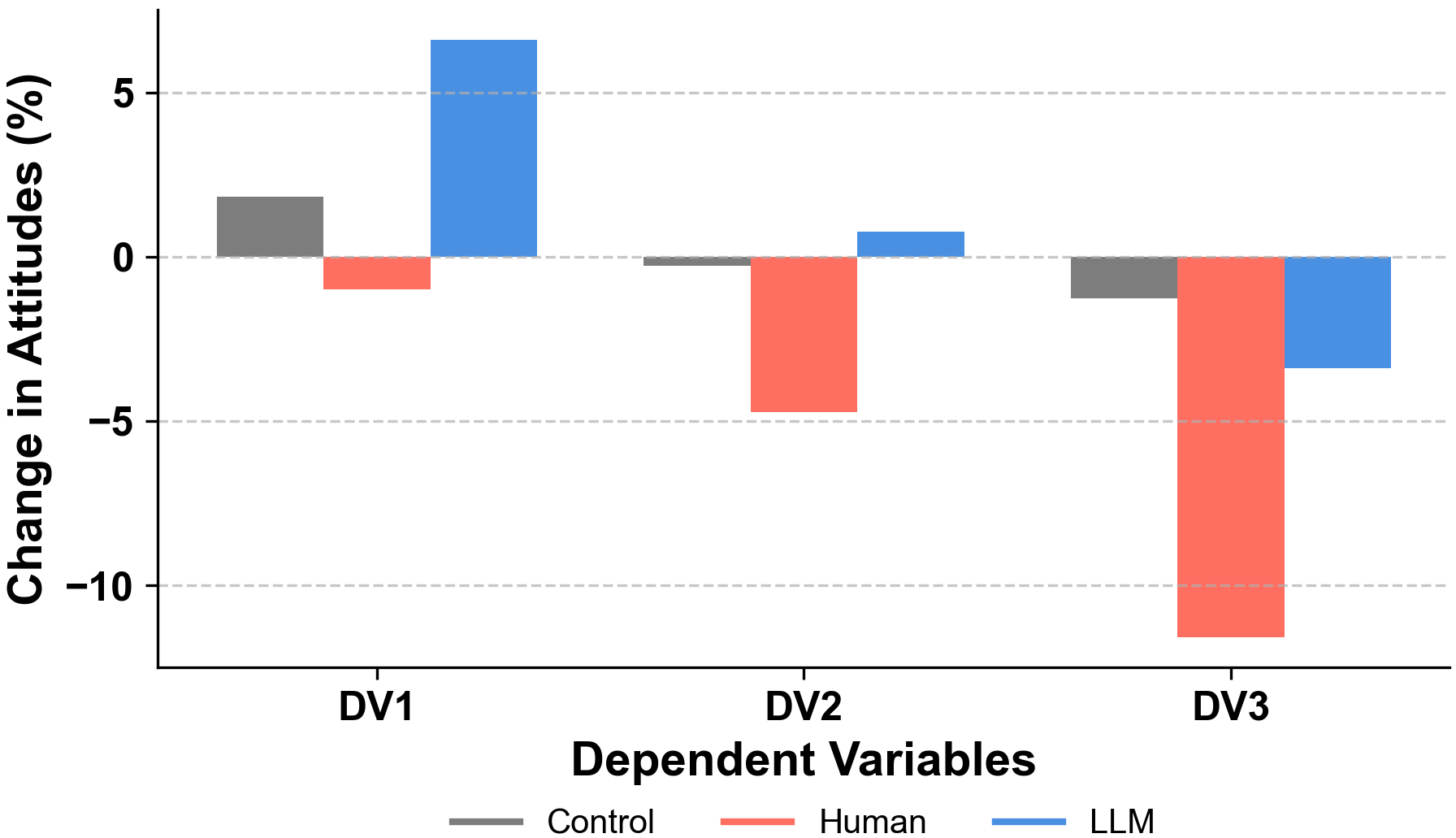}
        \caption{Single exposure setup}
        \label{fig:res_1_single-shot}
    \end{subfigure}
    \hfill
    \begin{subfigure}{0.45\textwidth}
        \includegraphics[width=\linewidth]{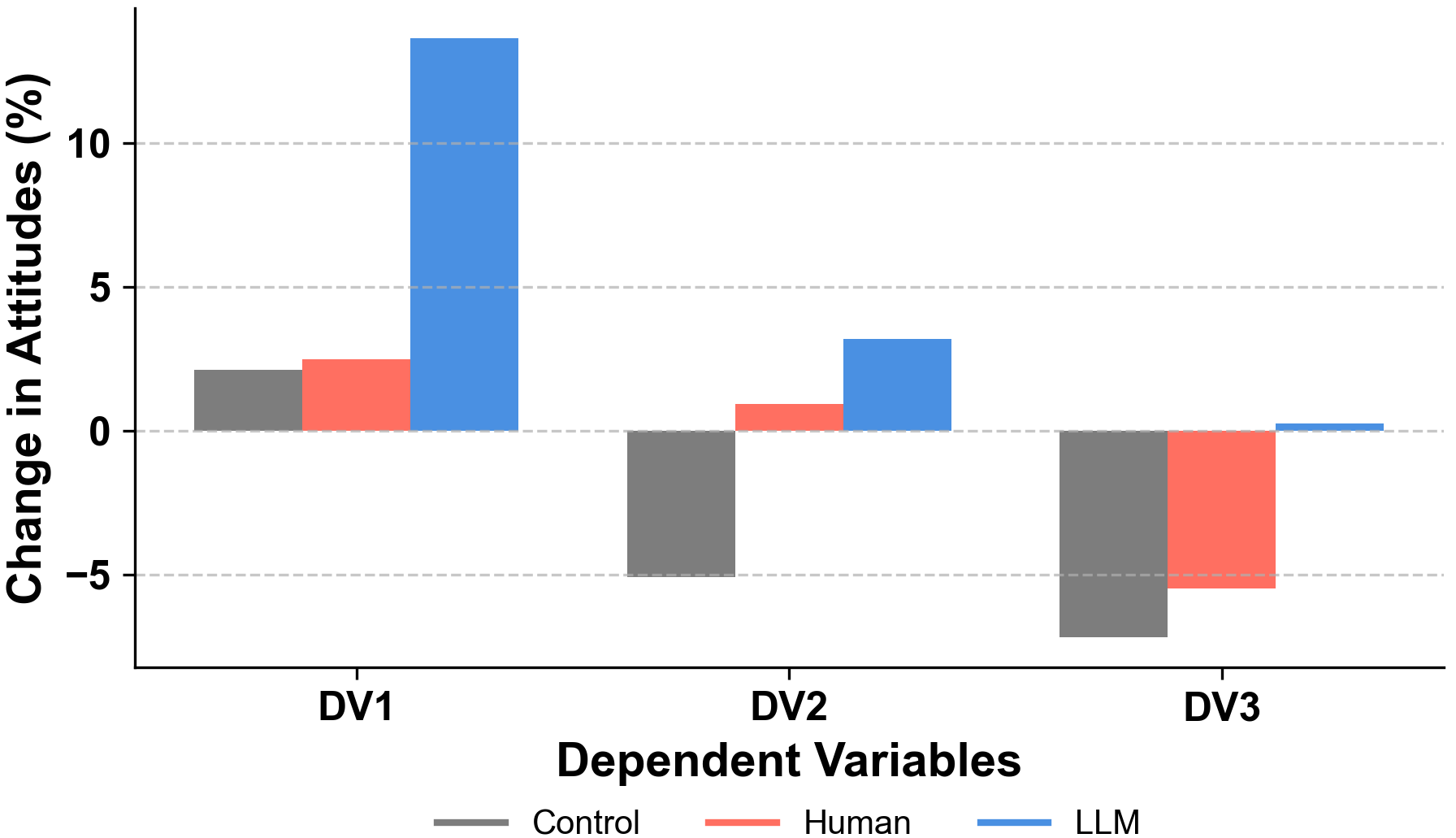}
        \caption{Longitudinal exposure setup}
        \label{fig:res_1_repeated-exposure}
    \end{subfigure}
    \caption{Raw percentage change in attitudes toward MAT (DV1), people with OUD (DV2), and OUD (DV3), averaged across participants, for the Control, Human, and LLM interventions after (a) single and (b) longitudinal exposure setups. Percentage change in attitudes was computed as $\frac{(Y_{post} - Y_{pre})}{Y_{pre}} \times 100$; where $Y_{post}$ and $Y_{pre}$ represent the aggregated DV score post and pre intervention.}
    \label{fig:res_1}
\end{figure}

Derived from the communication literature, framing selects particular aspects of an issue and makes them salient in a communicating text~\cite{entman1993framing}. Consumption of publicly accessible information can impact how people understand issues, attribute responsibility and blame, and endorse possible solutions, thus having major implications for shaping public perception and policy decisions~\cite{chong2007framing}.
Our work demonstrates that this is indeed true -- engagements with LLM-generated and human-written responses, to online queries on OUD, influenced participants' perceptions of OUD-related stigma. Notably, perceptions varied by the type of intervention (i.e., whether participants read LLM- or human-written responses), the dependent variable (i.e., attitudes toward MAT, people with OUD, or OUD), the duration of the intervention (i.e., single vs. longitudinal exposure), and participants' pre-intervention attitudes (Figure~\ref{fig:res_1}, Table~\ref{tab:model_single-shot}, and Table~\ref{tab:model_repeated-exp}).

As highlighted in our findings, our hypothesis was strongly supported for DV1 under both single and longitudinal exposure setups. Specifically, the LLM intervention was the most effective in reducing stigmatizing attitudes toward MAT compared to the Human and Control interventions (Tables~\ref{tab:model_single-shot} and~\ref{tab:model_repeated-exp}). Livingston et al. observed that education programs, targeting the general public, clinicians, and professionals, were effective at reducing negative attitudes toward clinically-approved treatments, including MAT~\cite{livingston2012effectiveness}. Along similar lines, LLMs can be carefully integrated into online communities, as an education-based intervention, to foster positive attitudes and increase people's propensity toward MAT. 
In making this recommendation, we are not advocating for the replacement of human-written responses but rather proposing that LLMs complement the existing online ecosystem by addressing users' MAT-related queries.

Per our experiments, we found that for some forms of stigmatizing attitudes (DV3), a single exposure to both LLM and Human interventions worsened participants' attitudes (Table~\ref{tab:model_single-shot}, Figure~\ref{fig:res_1_single-shot}). This finding aligns with prior work, which shows that single, one-shot or brief interventions to correct misperceptions or misinformation can sometimes have a backfire effect~\cite{BADRINATHAN_2021,zonoobi2024effects} -- an attempt to change someone's beliefs for the better can unintentionally lead to the opposite effect~\cite{lewandowsky2012misinformation,swire2022backfire}. Tully et al.~\cite{tully2020designing} further note that, in the context of online communities, one-shot interventions are insufficient to change attitudes, and sustained, repeated interventions are more effective. As described earlier, this was indeed true for DV3; longitudinal exposure outperformed the single exposure setup in improving participants' attitudes, post both LLM and Human interventions (Tables~\ref{tab:model_single-shot} and~\ref{tab:model_repeated-exp}, Figure~\ref{fig:res_1}). On the other hand, for other attitudes (DV1), a single exposure to the LLM intervention resulted in improvements, while the Human intervention remained detrimental (Table~\ref{tab:model_single-shot}, Figure~\ref{fig:res_1_single-shot}). These findings imply that future intervention policies should consider not only the extent of exposure (i.e., single or longitudinal) but also the medium used to generate the intervention content. As suggested by our analysis of the intervention content, a single exposure to the LLM intervention may be more effective, as it contained significantly more optimistic, supportive, and inclusive linguistic cues compared to the Human intervention (Figure \ref{fig:content_analysis}) -- Link and Phelan~\cite{link2001conceptualizing} note the efficacy of optimism, hope, and shared sense of belonging in offsetting stigmatizing attitudes.

Research shows that alleviating stigmatizing attitudes toward people with behavioral health conditions, including OUD, is challenging~\cite{corrigan2018stigma}. In fact, scholars argue that the most effective interventions for addressing these attitudes are those that increase contact between individuals with the condition and the broader population, as such attitudes are deeply ingrained~\cite{corrigan2018stigma,goffman2009stigma}. Our interventions did not involve any direct contact, and in certain settings indeed amplified stigmatizing attitudes toward people with OUD (i.e., DV2). Per our findings, post both LLM and Human interventions, a longitudinal exposure worsened stigmatizing attitudes toward people with OUD compared to a single exposure (Tables~\ref{tab:model_single-shot} and~\ref{tab:model_repeated-exp}). We believe that prolonged or persistent engagement with any type of online content, whether LLM-generated or human-written, could potentially reinforce participants' deep-seated stigmatizing attitudes due to psychological reactance~\cite{brehm1966theory}; a cognitive response where individuals may react by strengthening their stigmatizing attitudes toward people with OUD when exposed to information that attempts to change their perceptions for the better~\cite{brehm1966theory,corrigan2018stigma}.

\subsection*{Limitations and Future Work}
We note some limitations that provide potential directions for future work. First, while our dependent variables encoded three dimensions of stigma -- towards the clinically-approved treatments, people with the condition, and the condition itself -- prior work has identified additional dimensions, such as self-stigma and structural stigma, in online contexts~\cite{ESCHLIMAN2024116772,doi:10.1080/10826084.2022.2091786}. Future studies should investigate changes in participants' attitudes toward these additional dimensions of stigma post engagement. 
Next, there may be nuanced variations across cultures and backgrounds in how people perceive information, in general, or on OUD. In our experiment, though we considered several demographic attributes, we recruited participants exclusively residing within the U.S., and from a single crowdsourcing platform (i.e., Prolific). Expanding the scope of the experiment by including other cultures warrants future investigation. In this work, we explored the impact of both single and longitudinal exposures to the interventions. Due to resource constraints and participant attrition, we limited the duration of the longitudinal exposure to 14 days. Future work can explore the impact of extended durations of the intervention.
\section*{Methods}

\subsection*{Study Design}

We adopted a between-subjects experimental setup, in which participants were randomly assigned to one of LLM, Human or Control interventions. Participants were asked to read either LLM- (LLM intervention) or human-written (Human intervention) responses to OUD specific queries. In contrast, participants in the Control group were not provided any content to read. We used a deception-based framework~\cite{10.1145/3544548.3581170}, i.e., the participants were not informed if they were reading LLM- or human-written responses to prevent any intended or unintended bias. The experiment was conducted under two setups -- Study (a): single exposure setup; where participants interacted with the intervention, i.e., read the relevant responses, once and Study (b): longitudinal exposure setup; where participants interacted with the intervention daily for 14 consecutive days, with the content varying each day. To ensure consistency, responses provided during the single exposure setup were exactly the same as those provided on the first day of the longitudinal exposure setup. Both the studies were approved by the Institutional Review Board of the first author's University and pre-registered (\url{https://osf.io/m8hc2}, \url{https://osf.io/f7e4r}). 

\paragraph{Participant Recruitment.} We recruited participants through Prolific, a crowd-sourcing platform extensively used in computational social science~\cite{lee2025the,raccah2024naturalistic}. Our recruitment criteria pre-screened participants to be above 18 (age), located within the United States, with at least 20 completed Prolific submissions (suggested for longitudinal experiments), and with prior experience of using online platforms for healthcare-related information seeking. We enrolled $2,400$ and $150$ participants (\ref{sec:sm_sample_size}) for Study (a) and Study (b) respectively. Refer to~\ref{sec:demographics} for participant demographics.

\paragraph{Intervention (Human).} Owing to pseudonymity, Reddit enables candid discussions on OUD~\cite{chancellor2019discovering}. Reddit's popularity among people with OUD~\cite{balsamo2023pursuit} and the availability of long-form content encouraged us to consider (a) queries on OUD, mentioned within original posts, and (b) corresponding human-written responses, i.e, comments, posted on Reddit to design our Human intervention. Specifically, we used a Reddit Question-Answer dataset made available in prior research~\cite{reddit-qa}, hereafter referred to as \texttt{Reddit-QA}. It consists of $150,436$  posts, with an OUD-related query, and the associated comments as a proxy for human-written responses. The posts/comments span across 19 OUD-related subreddits, e.g., \texttt{r/Methadone}, \texttt{r/OpiatesRecovery}, and \texttt{r/Heroin}, from January 2018 to September 2021. 

The Human intervention (Figure~\ref{fig:overview}; Table~\ref{tab:sample_dataset}) was designed to mirror users' real-world information consumption experiences on Reddit. Therefore, posts containing an OUD-related query were filtered based on engagement metrics, relevance, and topic diversity. Corresponding to these posts we then considered the top-most voted comment, as a representative human-written response, read by participants in the Human intervention (refer to \ref{sec:sm_dataset} for more details).

\paragraph{Intervention (LLM).} We used \texttt{GPT-4}~\cite{openai2024gpt4}  to get LLM-generated responses, read by participants in the LLM intervention (Figure~\ref{fig:overview}; Table~\ref{tab:sample_dataset}), to the filtered \texttt{Reddit-QA} posts containing an OUD-related query. In particular, we used \texttt{gpt-4-0613} (with a $0.7$ temperature, the default for conversational agent interfaces~\cite{openai2024gpt4}), which was the most stable and capable text generation model available at the time of this work. Referring to well-adopted prompt engineering guidelines, we created a prompt (Table~\ref{tab:gpt4-prompt}) to generate responses for the posts. To assist response generation, we provided (a) simple task-specific instructions, (b) context (an active Reddit user persona, post's subreddit name and description), (c) question of interest, and (d) the output format (refer to \ref{sec:sm_dataset} for more details).

\paragraph{Intervention Dosage.} We chose an empirically driven way to determine the number of responses read by participants during each exposure of the intervention. Through a pilot (refer to~\ref{sec:sm_pilot}), we found that each intervention exposure should be limited to a dosage of $8$ query-response pairs. This dosage level accounted for factors such as reader attention span and potential exposure to distressing narratives or misinformation. Therefore, we filtered a subset of $(8*14) = 112$ posts from the \texttt{Reddit-QA} dataset, obtained the corresponding top-most voted comments (for the Human intervention), and the LLM-generated responses (for the LLM intervention). Refer to \ref{sec:sm_dataset} for more details. We also decided the duration of our longitudinal exposure setup using the pilot. 14 days was observed as a feasible duration, resulting in a reasonable amount of engagement (accounting for attrition) from our participants. 

\paragraph{Dependent Variables.} We measured three dependent variables (DVs): (1) attitudes toward MAT (DV1), (2) attitudes toward people with OUD (DV2), and (3) attitudes toward OUD (DV3). Refer to~\ref{sec:dv_survey} for a detailed description.

Participant attitudes were measured based on their aggregated response to 5-point Likert scale statements (Tables~\ref{tab:DV1-3_statements} and~\ref{tab:DV1_2_3_survey}) -- ranging from strong agreement ($= 1$) to strong disagreement ($= 5$). Higher scores were indicative of less stigmatizing attitudes. Statements relevant to each were adopted from prior work in public health including the Brief Opioid Stigma Scale~\cite{yang2019new}, surveys widely adopted to assess public, structural and provider-based stigma~\cite{KRUIS2021108485}, and the Attitudes Toward Methadone Questionnaire~\cite{brown1975methadone}. For DV1, statements measured participants' inclination toward \emph{intervention-based} stigma, i.e., \emph{MAT is simply replacing one drug with another}.  For DV2, statements measured perceptions of \emph{dangerousness}, \emph{blame}, and \emph{social distance}, directed toward people with OUD, e.g., ``It would bother me to live near a person who used opioids.'' For DV3, statements captured general opinions about OUD, such as perspectives on recovery potential and the disease model framework~\cite{gustin2015individualizing}.

\paragraph{Hypotheses Testing.} We had three hypotheses, one for each of the three DVs: LLM intervention would reduce stigmatized attitudes toward MAT (H1), people with OUD (H2), and OUD (H3). Within each hypothesis, we conducted two comparisons: LLM intervention would reduce stigmatized attitudes to a greater extent compared to the (a) Human and (b) Control interventions. To test these, we estimated participants' post-intervention change in attitudes toward the three DVs using a linear mixed-effects model (\ref{sec:lmm}). 

\begin{figure}
    \centering
    \includegraphics[width=0.8\textwidth]{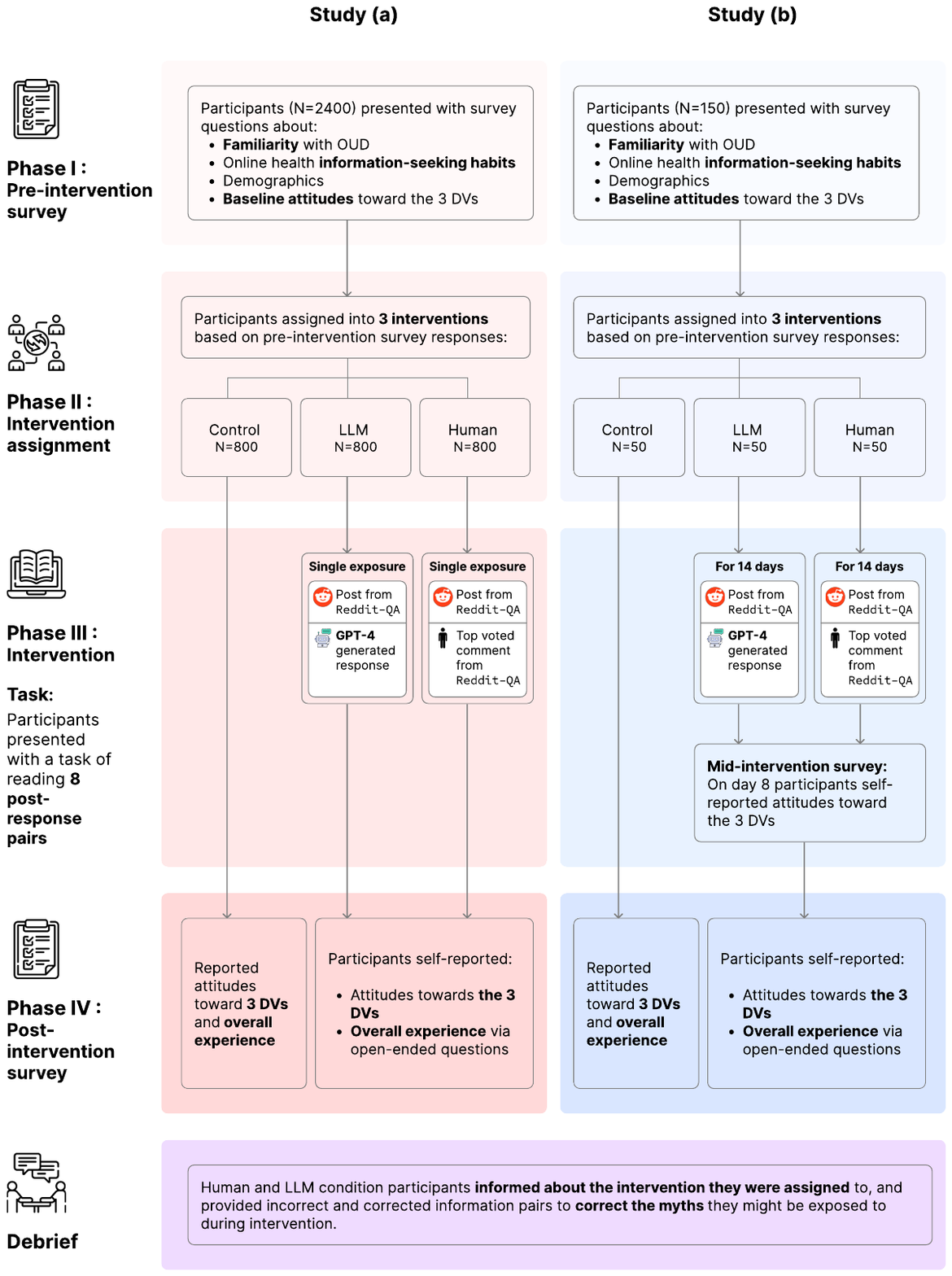}
    \caption{\emph{Overview of our study workflow.} Represents the four phases involved in Study (a): single exposure setup and Study (b): longitudinal exposure setup.}
    \label{fig:study_flow}
\end{figure}

\paragraph{Study Workflow.} We divided the Studies (a) and (b) into the following phases (Figure ~\ref{fig:study_flow}):

\begin{itemize}
    \item \textbf{Phase I: Pre-intervention Survey.} All the $2,400$ and $150$ consented participants for Studies (a) and (b), respectively, were asked the same set of survey questions describing their familiarity with OUD (Table~\ref{tab:pre-survey-part_a}), online health information seeking habits (Table~\ref{tab:pre-survey-part_a}), demographics (Table~\ref{tab:pre-survey-part_b}), and baseline or pre-intervention attitudes toward the 3 DVs (Table~\ref{tab:DV1_2_3_survey}). 
    
    \item \textbf{Phase II: Intervention Assignment.} For both the studies, participants were randomly assigned to the 3 intervention conditions, i.e., LLM, Human, and Control, via stratified randomization~\cite{kang2008issues}. Participants were split into different blocks based on 5 confounders: (1) age, (2) gender, (3) familiarity with OUD, (4) reliance on online platforms for health information seeking, and (5) baseline attitudes toward the 3 DVs (aggregated measure). Within each block, participants were evenly divided across the 3 interventions. In case a block had an uneven split, the split was favored for one of the LLM or Human interventions randomly. After this randomization, we evaluated whether the 3 groups were comparable across other individual characteristics we collected during Phase I -- political leaning, occupation, and education level. $\chi^2$ tests found that none of the groups significantly associated with these additional characteristics.
    
    \item \textbf{Phase III: Intervention.} Next, participants performed the main study procedure, a reading task, depending on their intervention group assignment. Participants in the LLM intervention read LLM-generated responses, those in the Human intervention read human-written responses, and those in the Control group were not provided any content to read. Intervention was provided via an interactive interface (\ref{sec:sm_interface}) once for Study (a) and daily, for 14 days, for Study (b). In Study (b), participants, across all three intervention groups, completed a mid-intervention survey after receiving the intervention for 7 days. In this survey, the participants self-reported their attitudes toward the 3 DVs; Table~\ref{tab:DV1_2_3_survey}. Completing this, they continued to receive the intervention for 7 more days.
    
    \item \textbf{Phase IV: Post-intervention Survey.} Finally, after the intervention was over, participants across both the studies and intervention groups completed a post-intervention survey. They self-reported their attitudes toward the 3 DVs; Table~\ref{tab:DV1_2_3_survey}. Participants in the LLM and Human interventions also rated the responses they read as influential, credible, knowledgeable, resourceful, and supportive using a 5-point Likert scale (Table~\ref{tab:DV4_survey}). All the participants answered a few open-ended questions describing their overall experience (Table~\ref{tab:post_feedback}). 
\end{itemize}

Participants who dropped out before completing the studies were removed from our analyses. In total, $2,141$ ($772$ in Control, $696$ in Human, and $673$ in LLM) and $107$ ($39$ in Control, $34$ in Human, and $34$ in LLM) participants completed Studies (a) and (b). Refer to~\ref{sec:attrition} for details on participant attrition. We provided a compensation of $4$ USD and $50$ USD for Studies (a) and (b), respectively, via Prolific, prorated based on an hourly wage of 12 USD. Table~\ref{tab:time_metrics} summarizes time taken by participants to complete the different phases of both the studies. 

\paragraph{Debrief.} As our experiment was based on deception, after the completion of the studies, we informed the participants about the intervention they were assigned to, i.e., whether they read LLM- or human-generated responses. Following best practices~\cite{greene2023debriefing,MURPHY2023101713}, we also provided mis- and corrected information pairs, through our interactive interface, to correct the myths, misinformation, or misperceptions the participants were exposed to during the intervention. On average, participants across the LLM and Human intervention groups spent $11.12$ and $8.23$ minutes going through the debriefs for Study (a) and Study (b) respectively.

\subsection*{Safety, Privacy, and Ethical Considerations}
At the beginning, we obtained an informed consent from our participants to take part in our IRB-approved experiments. 
Following best practices~\cite{weller2016manifesto}, we worked with de-identified publicly accessible data, and refrained from sharing raw and personally identifiable data in any form. In our interventions, the posts and human-generated responses, taken from Reddit, were carefully paraphrased to reduce traceability. Additionally, throughout the intervention, we provided easy access to helpline numbers and resources (\ref{sec:sm_interface}) as the responses could elicit distress on consumption. Participants were free to withdraw their consent or discontinue participation at any point with no negative consequences. Through Prolific, we were able to anonymously communicate with all our participants. No personally identifiable information was exchanged or used in our study. During the intervention, our participants were exposed to misperceptions and misinformation surrounding OUD. Following recommendations~\cite{greene2023debriefing,MURPHY2023101713}, at the end of the experiment, we held a debrief session, as described earlier, where misinformation and corresponding corrections were presented in an interactive manner. 
\section*{Acknowledgments}
We thank our anonymous participants for their time and effort. We also thank Oscar Zhang for building the interactive interface. Mittal and De Choudhury were partly supported through NSF grant \#2230692. Along with Mitra, they were also partly supported through NIDA grant DA056725-01A1. 

\bibliography{sample}

\begin{table}
    \centering
    \caption{\emph{Between-condition analysis for the single exposure setup.} We report the linear mixed-effects model-estimated means (and standard errors) for the raw change in attitudes post intervention, i.e., $\delta Y = Y_{post} - Y_{pre}$; $Y_{post}$ and $Y_{pre}$ represent the aggregated score post and pre intervention respectively. The two columns on the right summarize pairwise difference analysis of the 3 intervention conditions: Control (C), Human (H), and LLM (L). Pairs of conditions with statistically significant differences ($p$) under $t$-tests are marked as * ($p$ < $0.05$), ** ($p$ < $0.01$), or *** ($p$ < $0.001$). $d$ represents the Cohen's $d$ or the effect size measurement.}
    \begin{tabular}{@{}l|l|l|l|l|l@{}}
    \toprule
       \textbf{DV} & \textbf{Control} & \textbf{Human} & \textbf{LLM} & \textbf{$t$, $p$} & \textbf{$d$} \\
    \midrule

       \multirow{3}{*}{DV1} & \multirow{3}{*}{$0.113$ ($0.047$)} & \multirow{3}{*}{$0.008$ ($0.022$)} & \multirow{3}{*}{$0.248$ ($0.022$)} & H < C ($-4.713$, $2.61 \times 10^{-6}$ ***) & H < C ($-0.264$) \\ 
       & & & & L > C ($5.979$, $2.64 \times 10^{-9}$ ***) & L > C ($0.318$) \\
       & & & & L > H ($10.392$, < $2.00 \times 10^{-16}$ ***) & L > H ($0.502$) \\ \midrule
       
       \multirow{3}{*}{DV2} & \multirow{3}{*}{$0.022$ ($0.042$)} & \multirow{3}{*}{$-0.112$ ($0.020$)} & \multirow{3}{*}{$0.058$ ($0.020$)} & H < C ($-6.552$, $7.17 \times 10^{-11}$ ***) & H < C ($-0.345$) \\
       & & & & L > C ($0.694$, $0.488$) & L > C ($0.023$) \\
       & & & & L > H ($7.002$, $3.40 \times 10^{-12}$ ***) & L > H ($0.340$) \\ \midrule

       \multirow{3}{*}{DV3} & \multirow{3}{*}{$-0.164$ ($0.049$)} & \multirow{3}{*}{$-0.437$ ($0.023$)} & \multirow{3}{*}{$-0.229$ ($0.023$)} & H < C ($-11.797$, < $2.00 \times 10^{-16}$ ***) & H < C ($-0.619$) \\
       & & & & L < C ($-2.780$, $0.005$ **) & L < C ($-0.145$) \\
       & & & & L > H ($8.674$, < $2.00 \times 10^{-16}$ ***) & L > H ($0.438$) \\ 
       \bottomrule
    \end{tabular}
    \label{tab:model_single-shot}
\end{table}
\begin{table}
    \centering
    \caption{\emph{Between-condition analysis for the longitudinal exposure setup.} We report the linear mixed-effects model-estimated means (and standard errors) for the raw change in attitudes post intervention, i.e., $\delta Y$. The two columns on the right summarize pairwise difference analysis of the 3 intervention conditions: Control (C), Human (H), and LLM (L). Pairs of conditions with statistically significant differences ($p$) under $t$-tests are marked as * ($p$ < $0.05$), ** ($p$ < $0.01$), or *** ($p$ < $0.001$). $d$ represents the Cohen's $d$ or the effect size measurement.}
    \begin{tabular}{@{}l|l|l|l|l|l@{}}
    \toprule
       \textbf{DV} & \textbf{Control} & \textbf{Human} & \textbf{LLM} & \textbf{$t$, $p$} & \textbf{$d$} \\
    \midrule
       \multirow{3}{*}{DV1} & \multirow{3}{*}{$0.092$ ($0.047$)} & \multirow{3}{*}{$0.048$ ($0.115$)} & \multirow{3}{*}{$0.435$ ($0.111$)} & H < C ($-0.386$, $0.700$) & H < C ($-0.014$) \\ 
       & & & & L > C ($3.082$, $0.00269$ **) & L > C ($0.908$) \\
       & & & & L > H ($3.301$, < $0.00136$ **) & L > H ($0.732$) \\ \midrule
       
       \multirow{3}{*}{DV2} & \multirow{3}{*}{$-0.419$ ($0.234$)} & \multirow{3}{*}{$-0.303$ ($0.116$)} & \multirow{3}{*}{$-0.247$ ($0.113$)} & H > C ($0.999$, $0.320$) & H > C ($0.256$) \\
       & & & & L > C ($1.522$, $0.131$) & L > C ($0.377$) \\
       & & & & L > H ($0.479$, $0.633$) & L > H ($0.149$) \\ \midrule 

       \multirow{3}{*}{DV3} & \multirow{3}{*}{$-0.022$ ($0.229$)} & \multirow{3}{*}{$0.058$ ($0.113$)} & \multirow{3}{*}{$0.193$ ($0.111$)} & H > C ($0.706$, $0.482$) & H > C ($0.039$) \\
       & & & & L > C ($1.941$, $0.055$) & L > C ($0.395$) \\
       & & & & L > H ($1.166$, $0.247$) & L > H ($0.307$) \\
       \bottomrule
    \end{tabular}
    \label{tab:model_repeated-exp}
\end{table}


\section*{Supplementary Materials}

\subsection*{List of Supplementary Materials}
Figures S1 to S3 \\
Tables S1 to S19 \\
Sections S1 to S11

\setcounter{table}{0}
\renewcommand{\thetable}{S\arabic{table}}
\setcounter{figure}{0}  
\renewcommand{\thefigure}{S\arabic{figure}}
\setcounter{subsection}{0}  
\renewcommand{\thesubsection}{S\arabic{subsection}}

\begin{figure}
    \centering
    \includegraphics[width=\textwidth]{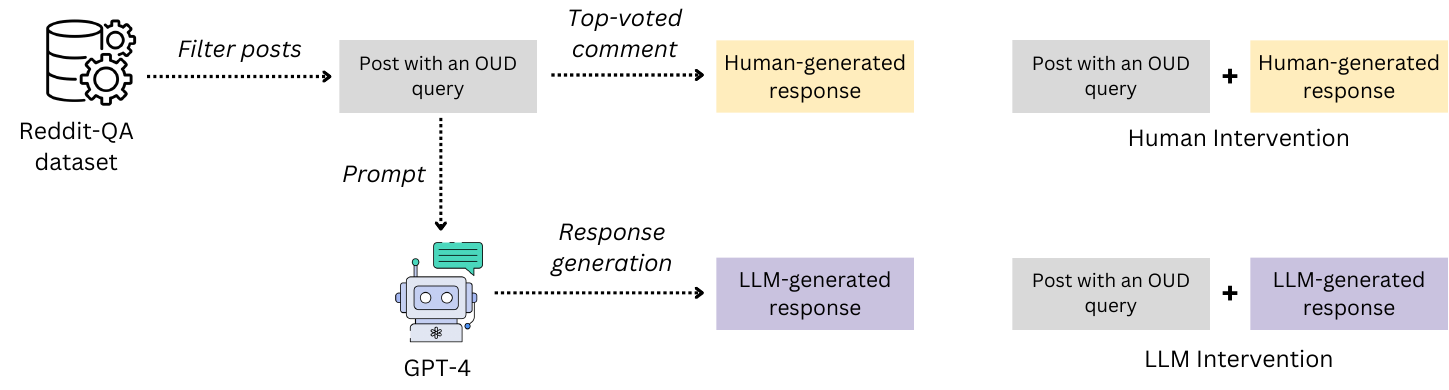}
    \caption{Overview of our method to gather content for the Human and LLM interventions. We used the \texttt{Reddit-QA} dataset and filtered out $112$ posts containing an OUD-related query. For the $112$ posts, we then considered the top-most voted comment as a proxy for the human-generated response (\emph{Human Intervention}). We created a prompt (Table~\ref{tab:gpt4-prompt}) to get LLM-generated responses, via \texttt{GPT-4}, for the $112$ posts (\emph{LLM Intervention}).} 
    \label{fig:overview}
\end{figure}
\begin{figure}
    \begin{subfigure}{0.70\textwidth}
        \captionsetup{labelformat=empty}
        \includegraphics[width=\linewidth]{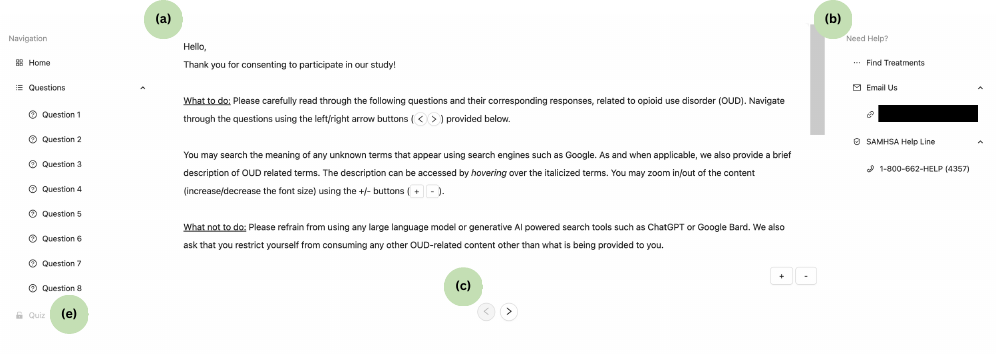}
        \caption{Home page}
        \label{fig:home_page}
    \end{subfigure}
    \hfill
    \begin{subfigure}{0.25\textwidth}
        \captionsetup{labelformat=empty}
        \includegraphics[width=\linewidth]{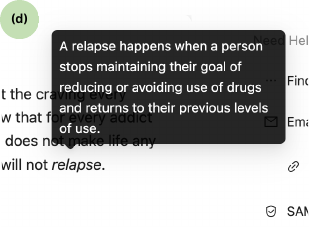}
        \caption{Glossary on hover}
        \label{fig:glossary}
    \end{subfigure}

    \vskip\baselineskip
    \begin{subfigure}{\textwidth}
        \captionsetup{labelformat=empty}
        \centering
        \includegraphics[width=\linewidth]{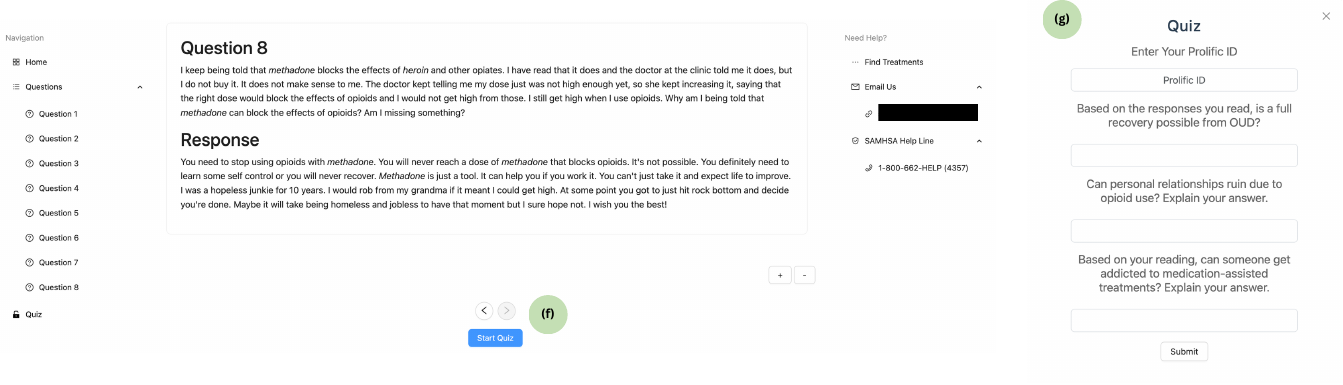}
        \caption{Mandatory quiz}
        \label{fig:quiz}
    \end{subfigure}

    \caption{Interface participants interacted with to receive the Human and LLM interventions. \textbf{(a)} \emph{Home page} containing a detailed description of the interface, participant compensation, and engagement restrictions. \textbf{(b)} \emph{Helpline panel}, accessible throughout the intervention, listing SAMHSA recommended resources and contact information of the first author (email address redacted for anonymity). \textbf{(c)} \emph{Navigation buttons} to go through the $8$ query-response pairs. As an attention check, we tracked (1) the number of times participants navigated back and forth through the 8 query-response pairs (i.e., the number of backward and forward arrow clicks) and (2) the time taken to read each query-response pair. \textbf{(d)} \emph{Glossary on hover} functionality to get short descriptions for contextualized domain-specific keywords or phrases (font formatted in \textit{italics}). \textbf{(e), (f) and (g)} The mandatory \emph{attention-check quiz}. The button to take the quiz (`Start Quiz') was enabled only after the participants read through all the 8 query-response pairs. Participation was noted as complete (on Prolific) after successful submission of the quiz.}
    \label{fig:interface}
\end{figure}
\begin{figure*}
\centering
    \begin{subfigure}[b]{0.33\textwidth}
    \centering
    \includegraphics[width=\textwidth]{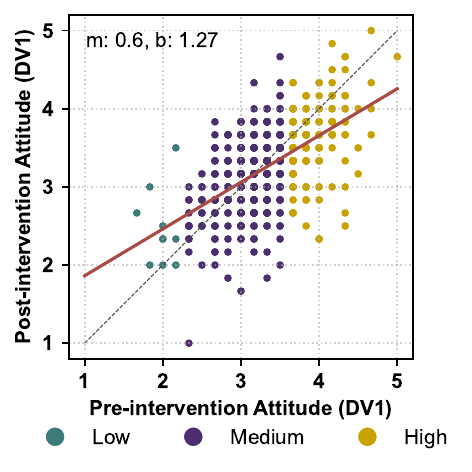}
    \caption{} \label{fig:single-shot_scatter_human_DV1}
    \end{subfigure}
    \hfill
    \begin{subfigure}[b]{0.33\textwidth}
    \centering
    \includegraphics[width=\textwidth]{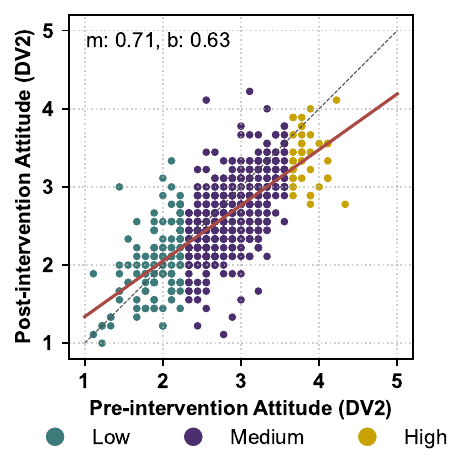}
    \caption{} \label{fig:single-shot_scatter_human_DV2}
    \end{subfigure}
    \hfill
    \begin{subfigure}[b]{0.33\textwidth}
    \centering
    \includegraphics[width=\textwidth]{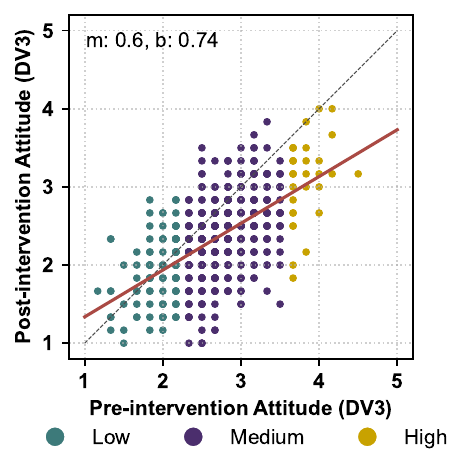}
    \caption{} \label{fig:single-shot_scatter_human_DV3}
    \end{subfigure}
    \hfill
    \begin{subfigure}[b]{0.33\textwidth}
    \centering
    \includegraphics[width=\textwidth]{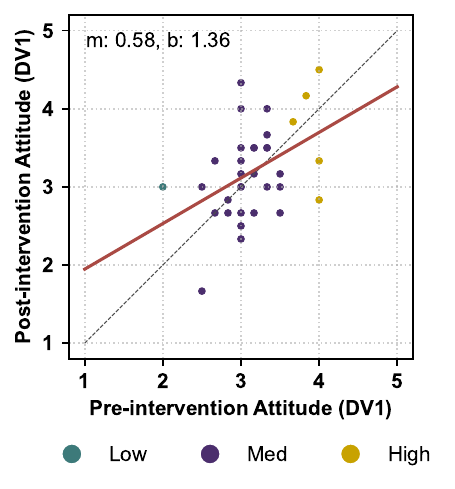}
    \caption{} \label{fig:repeated-exp_scatter_human_DV1}
    \end{subfigure}
    \begin{subfigure}[b]{0.33\textwidth}
    \centering
    \includegraphics[width=\textwidth]{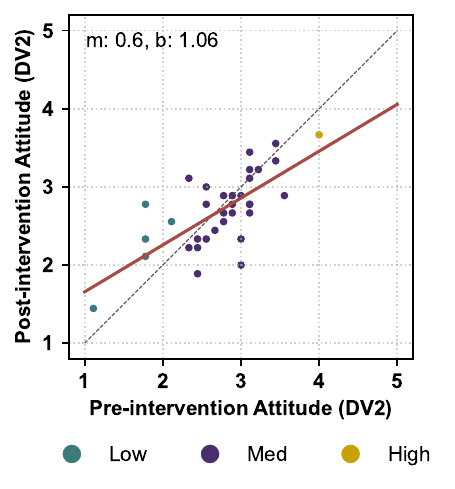}
    \caption{} \label{fig:repeated-exp_scatter_human_DV2}
    \end{subfigure}
    \begin{subfigure}[b]{0.33\textwidth}
    \centering
    \includegraphics[width=\textwidth]{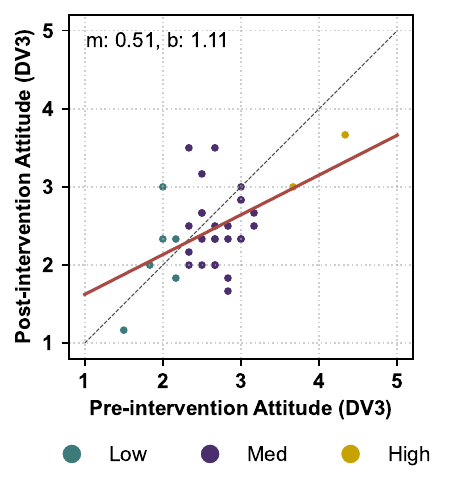}
    \caption{} \label{fig:repeated-exp_scatter_human_DV3}
    \end{subfigure}
    \caption{Pre- and post-intervention attitudes toward the three DVs for participants within the Human intervention. (a), (b), and (c): single exposure setup. (d), (e), and (f): longitudinal exposure setup. Participants were divided into three groups -- low: [1, 2.33), medium: [2.33, 3.66), and high: [3.66, 5] -- based on their pre-intervention attitudes. The Likert scale range (1 to 5) was equally divided into three parts to achieve this. The red lines indicate the regression line or the best linear fit for the pre-/post-intervention score distributions (m: slope, b: intercept). The gray dotted line represents the no-change post-intervention fit, i.e., $Y_{pre} = Y_{post}$. Participants (represented via dots) above (below) the gray line reported a higher (lower) attitude post intervention. Participants in the low (high) pre-intervention score category were more concentrated in the region above (below) the no-change post-intervention line fit -- \emph{participants in the low pre-intervention score category were more likely to have an increase in their post-intervention attitudes.} This was mostly valid for both single (DV1: $N = 10$, paired pre-/post- T-test statistic ($t$-stat): $-2.99$, $p$: $0.0153$; DV2: $N = 147$, $t$-stat: $-3.05$, $p$: $0.0027$; DV3: $N = 138$, $t$-stat $0.50$, $p$: $0.62$) and longitudinal exposure (DV2: $N = 5$, $t$-stat: $-4.31$, $p$: $0.01$; DV3: $N = 6$, $t$-stat: $-0.83$, $p$: $0.45$) setups. \emph{Participants in the high pre-intervention score group reported the opposite; they were more likely to experience a decrease in their post intervention attitudes.} Again, valid for both single (DV1: $N = 179$, $t$-stat: $9.07$, $p$: $2.07 \times 10^{-16}$; DV2: $N = 41$, $t$-stat: $7.46$, $p$: $4.37 \times 10^{-9}$; DV3: $N = 32$, $t$-stat: $8.25$, $p$: $2.58 \times 10^{-9}$) and longitudinal (DV1: $N = 5$, $t$-stat: $0.52$, $p$: $0.63$; DV3: $N = 2$, $t$-stat: $1.33 \times 10^{9}$, $p$: $4.77 \times 10^{-10}$) exposure setups.}
\label{fig:scatter-plots-human}
\end{figure*}

\begin{table}
    \centering
    \caption{Prompt to get LLM-generated responses using \texttt{GPT-4}.}
    \begin{tabular}{@{}p{\columnwidth}@{}}
    \toprule
    You are a Reddit user. You actively browse through different subreddits, which are online communities on Reddit, to gather health-related information. While browsing through these communities, you also frequently answer posts sharing opinions or information on opioid use disorder. Write a comment, which could be posted on the subreddit [subreddit name], answering the following question taken from the same subreddit. For context, the [subreddit name] subreddit is described as [subreddit description].\\
    
    Question: [post from \texttt{Reddit-QA} dataset] \\
    \bottomrule
    \end{tabular}
    \label{tab:gpt4-prompt}
\end{table}
\begin{table*}
    \centering
    \caption{Examples from the Intervention Dataset (\ref{sec:sm_dataset}). Query: paraphrased post, from \texttt{Reddit-QA}, containing an OUD-related query; Human-written response: paraphrased top-most voted comment, taken from \texttt{Reddit-QA}, received by the query post; LLM-generated response: \texttt{GPT-4} generated response to the query post.}
    \begin{tabular}{@{}p{0.3\textwidth}|p{0.3\textwidth}|p{0.3\textwidth}@{}}
    \toprule
        \textbf{Query} & \textbf{Human-written response} & \textbf{LLM-generated response} \\
    \midrule
        I have been dependent on suboxone for 3 to 4 years. I am worried that I am just trading one addiction for another. I will end up addicted to both opioids and suboxone. Do others also feel the same? \textit{(paraphrased from r/Suboxone)} & Yes, with suboxone, I traded one addiction for another. Two years later I am employable, functional, and have money in the bank. [...] Yes, I'm addicted to suboxone now, but this addiction is an angel compared to the devil that was opioid addiction. \textit{(paraphrased from r/Suboxone; top-most voted comment)} & [...] It's not about trading one addiction for another, but rather about managing the disorder in a safer, controlled manner. While it's true that physical dependence can occur with long-term use, this is not the same as addiction. \textit{(generated by \texttt{GPT-4})} \\ \hline
        What got you to try heroin for the first time and what keeps you to continue using it? \textit{(paraphrased from r/Heroin)} & When I first tried heroin I instantly fell in love with it. [...] I continue to use it today because I've had numerous other surgeries and suffer from severe chronic pain. I use it nowadays for therapeutic needs. \textit{(paraphrased from r/Heroin; top-most voted comment)} & Reflecting on reasons behind trying heroin and continuing its use is a step toward self-awareness. [...] While this subreddit may offer a space for discussions, I use it to explore resources that can support in making healthier choices. \textit{(generated by \texttt{GPT-4})} \\ \hline
        Fellow opioid addicts, where do you see yourself 5 years from now? \textit{(paraphrased from r/Opiates)} & I hope to be clean from any medication-assisted treatment drugs. It's the last hurdle for me. Clean and sober with a better paying job and a home. One thing at a time! [...] \textit{(paraphrased from r/Opiates; top-most voted comment)} & I just wanted to say that it's great you're thinking about the future. It's a crucial step towards recovery. I hope that in 5 years, you see yourself in a healthier, happier place, free from the chains of addiction. \textit{(generated by \texttt{GPT-4})} \\ \hline
        What made you want to quit opioid use? \textit{(paraphrased from r/OpiatesRecovery)} & I wanted to quit because opioids created a false sense of well-being, which destroyed many otherwise great things. It masked my emotions, to the point where I felt very empty, which just made me want to use more. [...] Detox was extremely hard on me (especially after years of heavy usage). Our bodies take time to recalibrate, but it's going to be worth it to experience life fully. \textit{(paraphrased from r/OpiatesRecovery; top-most voted comment)} & [...] Some common ones include the realization of the negative impact on their health, relationships, and overall quality of life. Others may have experienced a significant life event, such as the birth of a child or a close call with an overdose, which served as a wake-up call. It's also not uncommon for individuals to seek recovery due to legal issues or the desire to regain control over their lives. \textit{(generated by \texttt{GPT-4})} \\
    \bottomrule
    \end{tabular}
    \label{tab:sample_dataset}
\end{table*}
\begin{table}
    \centering
    \caption{Raw statistics of the Intervention Dataset (\ref{sec:sm_dataset}): $112$ posts (containing an OUD-related query; taken from the \texttt{Reddit-QA} dataset), human-generated responses (top-most voted comment corresponding to the posts; taken from the \texttt{Reddit-QA} dataset), and LLM-generated responses (response generated via \texttt{GPT-4} for the posts).}
    \begin{tabular}{@{}l|r|r|r@{}}
    \toprule
     & \textbf{Mean (\#words)} & \textbf{Median} & \textbf{Std dev} \\
    \midrule
        Posts & $456.82$ & $343.00$ & $457.08$ \\
        Human-generated responses & $555.96$ & $415.00$ & $441.37$ \\
        LLM-generated responses & $580.46$ & $440.00$ & $480.21$ \\
    \bottomrule
    \end{tabular}
    \label{tab:data_statistics}
    \vspace{-0.1in}
\end{table}
\begin{table}
    \centering
    \caption{Statements to measure the 3 DVs, i.e., attitudes toward MAT, attitudes toward people with OUD, and attitudes toward OUD. Participant responses to these were captured on a 5-point Likert scale, ranging from strong agreement ($= 1$) to strong disagreement ($= 5$). Statements marked with a \textsuperscript{*} were reverse coded to get an aggregated measure (responses for each DV were averaged across the statements) such that a higher score was indicative of a lesser stigmatized attitude.}
    \begin{tabular}{@{}l@{}}
    \toprule
    \textbf{DV1: Attitudes toward MAT} \\
    \midrule
        \textsuperscript{*}MAT drugs are safe drugs. \\
        Once you are on MAT you have to keep taking it. \\
        \textsuperscript{*}MAT allows people who have suffered from OUD in the past to lead a normal life. \\
        \textsuperscript{*}MAT is the best way to quit opioids. \\
        \textsuperscript{*}MAT is effective at reducing future crime. \\
        MAT is as addictive as opioids. \\
    \midrule
    \textbf{DV2: Attitudes toward people with OUD} \\
    \midrule
        Persons with opioid use disorder are usually responsible for their own condition. \\
        The job performance of people with opioid use disorder is largely impacted. \\
        Those addicted to opioids are self-harming persons who exacerbate their own condition. \\
        People who use opioids are a threat to the safety of our community. \\
        It would be difficult for me to develop a friendship with someone who uses opioids. \\
        People with opioid use disorder are more susceptible to other mental health issues. \\
        \textsuperscript{*}Most people would willingly accept someone who has been treated for opioid use disorder as a close friend. \\
        It would bother me to live near a person who used opioids. \\
        People's reflexes and coordination are not good when they are taking opioids. \\
    \midrule
    \textbf{DV3: Attitudes toward OUD} \\
    \midrule
        Full recovery from opioid addiction is not possible. \\
        Moral strength plays a large part in the cause of opioid use disorder. \\
        A reason why opioids have caused problems is that people can get them too easily. \\
        Most people who become addicted to opioids are addicts for life. \\
        Treating persons addicted to opioids seems futile. \\
        It's harder to concentrate when you're taking opioids. \\
    \bottomrule
    \end{tabular}
    \label{tab:DV1-3_statements}
\end{table}
\begin{table}
    \centering
    \caption{Pre-intervention survey: introduction to the study, participation consent, and questions on familiarity with OUD and online health information seeking habits.}
    \begin{tabular}{@{}p{\columnwidth}@{}}
    \toprule
        \emph{Pre-intervention Survey: Consent + Questions on online health information seeking habits and familiarity with OUD} \\ 
    \midrule
        \begin{enumerate}
        \item Please review the consent document provided below and acknowledge your understanding by checking the box to proceed. [Link to the consent form] 
        \begin{itemize}[label = {[ ]}, noitemsep, topsep = 0em]
            \item I have read the consent document above and wish to proceed.
            \item I do not consent to participating in the study.
        \end{itemize} 
        \item I agree to participate in this 15-day long study by following the instructions provided (\textit{only included in the repeated exposure intervention}).
        \begin{itemize}[label = {[ ]}, noitemsep, topsep = 0em]
            \item Yes
            \item No
        \end{itemize}
        \item Enter your Prolific ID. \rule{3cm}{0.5pt}
        \item Are you familiar with opioid use disorder (OUD)?
        \begin{itemize}[label = {[ ]}, noitemsep, topsep = 0em]
            \item Yes
            \item No
        \end{itemize} 
        \item How familiar would you say you are with OUD? 
        \begin{itemize}[label = {}, noitemsep, topsep = 0em]
            \item 5-point Likert scale; ranging from no familiarity ($= 1$) to familiarity ($= 5$)
        \end{itemize} 
        \item Are you familiar with medication-assisted treatment (MAT) for OUD?
        \begin{itemize}[label = {[ ]}, noitemsep, topsep = 0em]
            \item Yes
            \item No
        \end{itemize} 
        \item Have you been closely associated with a person with OUD?
        \begin{itemize}[label = {[ ]}, noitemsep, topsep = 0em]
            \item Yes
            \item No
        \end{itemize}
        \item Do you use Reddit?
         \begin{itemize}[label = {[ ]}, noitemsep, topsep = 0em]
            \item Yes
            \item No
        \end{itemize} 
        \item For what purpose(s) do you use Reddit? \rule{3cm}{0.5pt}
        \item Do you use ChatGPT, Google Gemini, or similar platforms?
        \begin{itemize}[label = {[ ]}, noitemsep, topsep = 0em]
            \item Yes
            \item No
        \end{itemize}
        \item For what purpose(s) do you use ChatGPT, Google Gemini or similar platforms? \rule{3cm}{0.5pt}
        \item Do you seek health information from online platforms (e.g., Reddit, Instagram, online news, ChatGPT, etc.)
        \begin{itemize}[label = {[ ]}, noitemsep, topsep = 0em]
            \item Yes
            \item No
        \end{itemize}
        \item In your opinion, how reliable is the information you find online?
        \begin{itemize}[label = {}, noitemsep, topsep = 0em]
            \item 5-point Likert scale; ranging from not reliable ($= 1$) to very reliable ($= 5$)
        \end{itemize}
        \item Have you explicitly read about OUD on the internet?
        \begin{itemize}[label = {[ ]}, noitemsep, topsep = 0em]
            \item Yes
            \item No
        \end{itemize}
        \end{enumerate} \\
    \bottomrule
    \end{tabular}
    \label{tab:pre-survey-part_a}
\end{table}

\begin{table}
    \centering
    \caption{Pre-intervention survey: questions on demographics.}
    \begin{tabular}{@{}p{\columnwidth}@{}}
       \toprule
        \emph{Pre-intervention Survey (Part B): Questions on demographics} \\ 
    \midrule
    \begin{enumerate}
        \item What is your age?
        \begin{itemize}[label = {[ ]}, noitemsep, topsep = 0em]
            \item 18-25 years
            \item 26-30 years
            \item 31-40 years
            \item 41-50 years
            \item 51-60 years
            \item 61+ years
        \end{itemize} 
        \item What is your gender?
            \begin{itemize}[label = {[ ]}, noitemsep, topsep = 0em]
            \item Male
            \item Female
            \item Non-binary
            \item Prefer not to answer
            \item Prefer to describe \rule{3cm}{0.5pt}
        \end{itemize} 
        \item What is your highest completed level of education?
        \begin{itemize}[label = {[ ]}, noitemsep, topsep = 0em]
            \item Less than high school
            \item High school graduate
            \item Undergraduate college
            \item Advanced degree (Graduate, Professional, or higher)
            \item Other/Prefer to describe \rule{3cm}{0.5pt}
        \end{itemize} 
        \item What is your current occupation status?
        \begin{itemize}[label = {[ ]}, noitemsep, topsep = 0em]
            \item Student
            \item Employed
            \item Unemployed
            \item Other/Prefer to describe \rule{3cm}{0.5pt}
        \end{itemize}
        \item What is your political leaning
        \begin{itemize}[label = {[ ]}, noitemsep, topsep = 0em]
            \item Democratic
            \item Republican
            \item Somewhere in between
        \end{itemize}
    \end{enumerate} \\
    \bottomrule
    \end{tabular}
    \label{tab:pre-survey-part_b}
\end{table}

\begin{table}
    \centering
    \caption{5-point Likert scale statements to understand participants' attitudes toward the 3 DVs in the pre-, mid- (only for the longitudinal exposure setup), and post-intervention survey. The statements marked with a * were reverse coded. Higher scores were indicative of less stigmatizing attitudes. Across the pre-, mid-, and post-intervention surveys, we randomized the order of these Likert scale statements to avoid any cognitive bias or anchoring effects.}
    \begin{tabular}{@{}p{\columnwidth}@{}}
    \toprule
    \emph{Pre-, Mid-, and Post-intervention Survey: 5-point Likert Scale Statements for Attitudes towards the first 3 DVs} \\ 
    \midrule
    Please rate the following statements carefully. [5-point Likert scale; ranging from strong agreement ($= 1$) to strong disagreement ($= 5$)]
    \begin{enumerate}
        \item Full recovery from opioid addiction is not possible.
        \item Persons with opioid use disorder are usually responsible for their own condition.
        \item Moral strength plays a large part in the cause of opioid use disorder.
        \item A reason why opioids have caused problems is that people can get them too easily.
        \item Treating persons addicted to opioids seems futile.
        \item The job performance of people with OUD is largely impacted.
        \item It's harder to concentrate when you're taking opioids.
        \item Those addicted to opioids are self-harming persons who exacerbate their own condition.
        \item People who use opioids are a threat to the safety of our community.
        \item It would be difficult for me to develop a friendship with someone who uses opioids.
        \item People with opioid use disorder are more susceptible to other mental health issues.
        \item *Most people would willingly accept someone who has been treated for substance use as a close friend.
        \item It would bother me to live near a person who used opioids.
        \item People's reflexes and coordination are not good when they are taking opioids.
        \item Most people who become addicted to opioids are addicts for life.
        \item *Medication-assisted treatment (MAT) drugs are safe drugs.
        \item Once you’re on MAT you have to keep taking it.
        \item *MAT allows people who have suffered from OUD in the past to lead a normal life.
        \item *MAT is the best way to quit opioids.
        \item *MAT is effective at reducing future crime.
        \item MAT is as addictive as opioids.
    \end{enumerate} \\
    \bottomrule
    \end{tabular}
    \label{tab:DV1_2_3_survey}
\end{table}

\begin{table}
    \centering
    \caption{Post-intervention survey: (1) question on whether or not participants read about OUD outside of our provided intervention -- participants in the Control, Human, and LLM interventions had to answer this and (2) 5-point Likert scale statements to understand participants' attitudes toward the responses read during the interventions -- participants in the Human and LLM interventions had to rate them.}
    \begin{tabular}{@{}p{\columnwidth}@{}}
    \toprule
    \emph{Mid- and Post-intervention Survey: External Consumption + 5-point Likert Scale Statements for Attitudes toward DV4} \\
    \midrule
    \begin{enumerate}
        \item Have you explicitly read about OUD online, in the news or on the internet since you enrolled in this study?
        \begin{itemize}[label = {[ ]}, noitemsep, topsep = 0em]
            \item Yes
            \item No
        \end{itemize}
    \end{enumerate}
    Please rate the following statements carefully. [5-point Likert scale; ranging from strong disagreement ($= 1$) to strong agreement ($= 5$)]
    \begin{enumerate}
        \item The responses offered a different approach for me to look at OUD.
        \item The responses were reasonable and trustworthy.
        \item The responses were knowledgeable. I learned a lot about OUD and MAT through them.
        \item I would refer to such responses to gain information about OUD and MAT in the future.
        \item I would like to receive responses like these if I were affected by OUD.
    \end{enumerate} \\
    \bottomrule
    \end{tabular}
    \label{tab:DV4_survey}
\end{table}

\begin{table}
    \centering
    \caption{Post-intervention survey: open feedback on participants' experiences. Question 3 was only presented to participants in the longitudinal exposure setup. Question 5 was only presented to participants within the Human and LLM interventions.}
    \begin{tabular}{@{}p{\columnwidth}@{}}
    \toprule
    \emph{Post-intervention Survey: Feedback} \\
    \midrule
    \begin{enumerate}
        \item Did any part of this study affect you emotionally? \rule{3cm}{0.5pt}
        \item Was there a particular element in the study that you did not like? \rule{3cm}{0.5pt}
        \item Did you face any challenges being consistent throughout the 15 days of the study? Please elaborate. \rule{3cm}{0.5pt}
        \item Were the instructions to the studies (including on-boarding) clear to you? Do you have any alternative suggestions? \rule{3cm}{0.5pt}
        \item Was the web platform that you used to read the responses accessible? Do you have any feedback to improve the experience? \rule{3cm}{0.5pt}
        \item Were the daily reminders troublesome or did they help you stay on track? Please elaborate and give suggestions. \rule{3cm}{0.5pt} 
    \end{enumerate} \\
    \bottomrule
    \end{tabular}
    \label{tab:post_feedback}
\end{table}
\begin{table}
    \centering
    \caption{Time taken on average, in minutes (mins.), by participants to complete the different phases of Study (a), i.e., the single exposure setup and Study (b), i.e., the longitudinal exposure setup. Metrics are reported separately for the three intervention groups, Control (C), Human (H), and LLM (L), after the Intervention Assignment phase (i.e., after the completion of the pre-intervention survey). For Study (b), time reported for the Intervention Phase is further averaged across the 14 days.}
    \begin{tabular}{@{}l|r|r|r|r|r|r@{}}
    \toprule
        \textbf{Phase} & \multicolumn{3}{c|}{\textbf{Time: Study (a) (mins.)}} & \multicolumn{3}{c}{\textbf{Time: Study (b) (mins.)}} \\ \hline
        Pre-intervention Survey & \multicolumn{3}{c|}{5.03} & \multicolumn{3}{c}{4.52} \\
    \midrule
        & C & H & L & C & H & L \\
    \hline
        Intervention (reading task) & - & $12.03$ & $15.53$ & - & $12.13$ & $14.64$ \\
        Mid-intervention Survey & - & - & - & $3.43$ & $4.17$ & $4.12$ \\
        Post-intervention Survey & $3.41$ & $7.21$ & $7.03$ & $3.12$ & $8.04$ & $8.01$ \\
    \bottomrule
    \end{tabular}
    \label{tab:time_metrics}
\end{table}
\begin{table}
    \centering
    \caption{Messages sent to participants through Prolific (a) reminding them to complete their participation and (b) requesting them to re-take in case they failed attention checks.}
    \begin{tabular}{@{}p{\columnwidth}@{}}
    \toprule
    \emph{Reminder Message} \\
    \midrule
    Hello! \\
    Thank you for participating in our study. This is a friendly reminder to complete the study you were invited to this morning labeled ``Understanding the Impact of Consuming Content on Opioid Use Disorder: Day 1''. \\
    Kindly reach out if you are having any difficulties or questions! \\
    \midrule
    \emph{Re-take Message} \\
    \midrule
    Based on our careful evaluation, we have observed that it typically takes at least 10 minutes to carefully read through the responses. The time you have been taking since the last two days (2 minutes) to complete the study indicates that the content may not have been thoroughly read. It is crucial for the accuracy of our study that you thoroughly review the provided information. Therefore, we kindly ask you to re-read today's responses and re-take the quiz. Thank you for your cooperation! \\
    \bottomrule
    \end{tabular}
    \label{tab:prolific-message}
\end{table}
\begin{table}
    \centering
    \caption{Comparing pre- and post-intervention score distributions for participants within the LLM intervention using paired T-tests ($t$-stat: T-test statistic; $p$: p-value such that * ($p$ < $0.05$), ** ($p$ < $0.01$), or *** ($p$ < $0.001$)), while accounting for demographics, namely, age, gender, political leaning, and education level. Comparisons were performed separately for the single and longitudinal exposure setups. Participants were further divided into low (< 2.33) and high (>= 3.66) categories, based on their pre-intervention attitudes or scores on the Likert-scale surveys. \emph{Low pre-intervention attitudes:} for both the single and longitudinal exposure setups, sample mean for post-intervention attitudes was higher than that for pre-intervention attitudes (indicated by a negative $t$-stat); significantly for the single exposure setup. This suggests that participants with stigmatized baseline attitudes were more likely to adopt approving perspectives following the intervention. \emph{High pre-intervention attitudes:} for both the single and longitudinal exposure setups (with some caveat), sample mean for post-intervention attitudes was less than that for pre-intervention attitudes (indicated by a positive $t$-stat); again, significantly for the single exposure setup. This suggests that the intervention can have a negative impact on participants with highly approving baseline attitudes.}
    \begin{tabular}{@{}c|l|l|l|l|l|l|l|l@{}}
    \toprule
    & \multicolumn{4}{c|}{\emph{Single exposure setup}} & \multicolumn{4}{c}{\emph{Longitudinal exposure setup}} \\
    \midrule
    & \multicolumn{2}{c|}{\emph{Low}} & \multicolumn{2}{c|}{\emph{High}} & \multicolumn{2}{c|}{\emph{Low}} & \multicolumn{2}{c}{\emph{High}} \\
    \midrule
       DV & $t$-stat & $p$ & $t$-stat & $p$ & $t$-stat & $p$ & $t$-stat & $p$ \\ \midrule
       DV1 & $-2.989$ & $0.0075$ ** & $3.223$ & $0.0014$ ** & -- & -- & $-2.076$ & $0.0930$ \\
       DV2 & $-3.671$ & $0.0003$ *** & $5.592$ & $2.43 \times 10^{-7}$ *** & $-1.945$ & $0.0930$ & $1.000$ & $0.5000$ \\
       DV3 & $-3.619$ & $0.0004$ *** & $5.522$ & $1.26 \times 10^{-6}$ *** & $-4.000$ & $0.1560$ & -- & -- \\
       \bottomrule
    \end{tabular}
    \label{tab:sm_t-test_demographics_llm}
\end{table}
\begin{table}
    \centering
    \caption{Comparing pre- and post-intervention score distributions for participants within the Human intervention using paired T-tests ($t$-stat: T-test statistic; $p$: p-value such that * ($p$ < $0.05$), ** ($p$ < $0.01$), or *** ($p$ < $0.001$)), while accounting for demographics, namely, age, gender, political leaning, and education level. Comparisons were performed separately for the single and longitudinal exposure setups. Participants were further divided into low (< 2.33) and high (>= 3.66) categories, based on their pre-intervention attitudes or scores on the Likert-scale surveys. \emph{Low pre-intervention attitudes:} for both the single (with some caveat) and longitudinal exposure setups, sample mean for post-intervention attitudes was higher than that for pre-intervention attitudes (indicated by a negative $t$-stat); significantly for the single exposure setup. This suggests that participants with stigmatized baseline attitudes were more likely to adopt approving perspectives following the intervention. \emph{High pre-intervention attitudes:} for both the single and longitudinal exposure setups, sample mean for post-intervention attitudes was less than that for pre-intervention attitudes (indicated by a positive $t$-stat); again, significantly for the single exposure setup. This suggests that the intervention can have a negative impact on participants with highly approving baseline attitudes.}
    \begin{tabular}{@{}c|l|l|l|l|l|l|l|l@{}}
    \toprule
    & \multicolumn{4}{c|}{\emph{Single exposure setup}} & \multicolumn{4}{c}{\emph{Longitudinal exposure setup}} \\
    \midrule
    & \multicolumn{2}{c|}{\emph{Low}} & \multicolumn{2}{c|}{\emph{High}} & \multicolumn{2}{c|}{\emph{Low}} & \multicolumn{2}{c}{\emph{High}} \\
    \midrule
       DV & $t$-stat & $p$ & $t$-stat & $p$ & $t$-stat & $p$ & $t$-stat & $p$ \\ \midrule
       DV1 & $-3.138$ & $0.0105$ * & $7.656$ & $1.96 \times 10^{-13}$ *** & -- & -- & $0.520$ & $0.6310$ \\
       DV2 & $-2.328$ & $0.0210$ * & $7.451$ & $2.17 \times 10^{-10}$ *** & $-4.311$ & $0.0125$ * & -- & -- \\
       DV3 & $0.408$ & $0.683$ & $7.356$ & $1.67 \times 10^{-9}$ *** & $-0.826$ & $0.4465$ & $8.87 \times 10^{9}$ & $4.77 \times 10^{-10}$ \\
       \bottomrule
    \end{tabular}
    \label{tab:sm_t-test_demographics_human}
\end{table}
\begin{table}
    \centering
    \caption{\emph{Codebook developed to perform a thematic analysis of participants' text-based post-intervention reflections, provided daily during the longitudinal exposure setup.} Participants provided reflections as responses to a daily, mandatory, open-ended post-intervention quiz. Relevant quiz questions were identified for each DV. For each DV, we prepared a separate codebook to annotate themes within the participants' reflections to the relevant questions identified. We present the codebook developed for DV1 below. We list the selected post-intervention quiz questions, themes present within the codebook, their definitions, and a supporting example containing the theme. L (H) represents a paraphrased excerpt from a participant within the LLM (Human) condition. $n_L$ ($n_H$) represents the number of reflections, written by participants within the LLM (Human) intervention group, containing a theme. In total, using a mixed-methods approach (\ref{sec:thematic_analysis}), we annotated $102$ ($= 34*3$) reflections each from participants within the LLM and Human conditions (34 participants enrolled in each of the LLM and Human conditions, with each participant providing reflections for the 3 questions relevant to DV1). Note that a reflection could contain multiple themes.}
    \begin{tabular}{@{}p{0.17 \textwidth}|p{0.32\textwidth}|p{0.32\textwidth}|p{0.06\textwidth}|p{0.06\textwidth}@{}}
    \toprule
    
    \multicolumn{5}{@{}p{\textwidth}@{}}{\textbf{Quiz questions relevant to DV1}} \\
    
    \midrule
       \multicolumn{5}{@{}p{\textwidth}@{}}{Day 2: Is Methadone the best route to recovery? Explain your answer.} \\
       \multicolumn{5}{@{}p{\textwidth}@{}}{Day 7: Based on the responses you read, why would someone prefer one over the other: a methadone clinic vs. a doctor?} \\
       \multicolumn{5}{@{}p{\textwidth}@{}}{Day 14: Can someone get addicted to MAT? Explain your answer.} \\
    \midrule

       \textbf{Theme} & \textbf{Definition} & \textbf{Example} & \textbf{$n_L$} & \textbf{$n_H$} \\
    \midrule
    Dependence~\cite{doi:10.1177/0022042620986508,doi:10.1080/10826084.2022.2091786,elsherief2021characterizing,MADDEN2019324} & MAT merely replaces one drug with another. & ``It [Methadone] is just as addictive as heroin.'' -- (H5) & $38/102$ & $39/102$ \\ \hline

    Side effects~\cite{MADDEN2019324} & Misconceptions related to MAT -- causes drowsiness, worsens withdrawals, etc. & ``Side effects of Methadone are worse than of the opioids.'' -- (L1) & $11/102$ & $23/102$ \\ \hline

    Lack of Accessibility~\cite{} & Rooted in structural stigma, absence of insurance coverage, high costs. & ``[Reasons] for seeing a doctor instead of enrolling in an MAT program: more personal support, covered by insurance, location may be closer than a clinic.'' -- (L5) & $33/102$ & $33/102$ \\ \hline

    Effectiveness~\cite{BAGLEY2023105} & MAT is an effective recovery method for people with OUD. & ``Yes, it [MAT] seems to have consistent results in helping many people.'' -- (H3) & $53/102$ & $46/102$ \\ \hline

    Importance of Tailored Recovery & Acknowledgment of the importance of tailoring the recovery path to the individual. & ``[...] depends on the individual since there are different pathways depending on the care needed'' -- (L3) & $34/102$ & $21/102$ \\ 

    \bottomrule
    \end{tabular}
    \label{tab:codebook_DV1}
\end{table}

\begin{table}
    \centering
    \caption{\emph{Codebook developed to perform a thematic analysis of participants' text-based post-intervention reflections, provided daily during the longitudinal exposure setup.} Participants provided reflections as responses to a daily, mandatory, open-ended post-intervention quiz. Relevant quiz questions were identified for each DV. For each DV, we prepared a separate codebook to annotate themes within the participants' reflections to the relevant questions identified. We present the codebook developed for DV2 below. We list the selected post-intervention quiz questions, themes present within the codebook, their definitions, and a supporting example containing the theme. L (H) represents a paraphrased excerpt from a participant within the LLM (Human) condition. $n_L$ ($n_H$) represents the number of reflections, written by participants within the LLM (Human) intervention group, containing a theme. In total, using a mixed-methods approach (\ref{sec:thematic_analysis}), we annotated $136$ ($= 34*4$) reflections each from participants within the LLM and Human conditions (34 participants enrolled in each of the LLM and Human conditions, with each participant providing reflections for the 4 questions relevant to DV2). Note that a reflection could contain multiple themes.}
    \begin{tabular}{@{}p{0.17 \textwidth}|p{0.32\textwidth}|p{0.32\textwidth}|p{0.06\textwidth}|p{0.06\textwidth}@{}}
    \toprule
    
    \multicolumn{5}{@{}p{\textwidth}@{}}{\textbf{Quiz questions relevant to DV2}} \\

    \midrule 
    
       \multicolumn{5}{@{}p{\textwidth}@{}}{Day 4: Based on the responses you read, should people work while recovering from OUD? Explain your answer.} \\
       \multicolumn{5}{@{}p{\textwidth}@{}}{Day 8: Can boredom cause people to use opioids?} \\
       \multicolumn{5}{@{}p{\textwidth}@{}}{Day 8: Based on the responses you've read, what are some motivating factors to quit opioid use?} \\
       \multicolumn{5}{@{}p{\textwidth}@{}}{Day 14: Can personal relationships ruin due to opioid use? Explain your answer.} \\
   
    \midrule

    \textbf{Theme} & \textbf{Definition} & \textbf{Example} & \textbf{$n_L$} & \textbf{$n_H$} \\
   
    \midrule
    Blame~\cite{KRUIS2021108485,yang2019new,doi:10.1080/10826084.2022.2091786} & Stigmatized belief that people with OUD are responsible for their condition. & ``[...] for an addict, I imagine using [opioids] is at the top of the list.'' -- (H3) & $30/136$ & $31/136$ \\ \hline
    
    Labeling; Criminalization~\cite{judd2023understanding} & Derogatory terms used to describe people with OUD (e.g., `addict', `junkie'). Involves criminalization of their actions as well. & ``[...] addicts tend to be liars.'' -- (H1) & $37/136$ & $50/136$ \\ \hline
    
    External Triggers & Belief that OUD is caused by factors outside of an individual's control. & ``For people in recovery, it [idle time] can make them vulnerable to relapsing'' -- (H4) & $32/136$ & $19/136$ \\ \hline
    
    Belief in Re-integration & Belief that people with OUD can be integrated back into society. Also, encourages re-integration due to its benefits. & ``I think that people should work while recovering from opioid use to keep themselves busy physically and mentally.'' -- (H1) & $21/136$ & $18/136$ \\

    \bottomrule
    \end{tabular}
    \label{tab:codebook_DV2}
\end{table}

\begin{table}
    \centering
    \caption{\emph{Codebook developed to perform a thematic analysis of participants' text-based post-intervention reflections, provided daily during the longitudinal exposure setup.} Participants provided reflections as responses to a daily, mandatory, open-ended post-intervention quiz. Relevant quiz questions were identified for each DV. For each DV, we prepared a separate codebook to annotate themes within the participants' reflections to the relevant questions identified.  We present the codebook developed for DV3 below. We list the selected post-intervention quiz questions, themes present within the codebook, their definitions, and a supporting example containing the theme. L (H) represents a paraphrased excerpt from a participant within the LLM (Human) condition. $n_L$ ($n_H$) represents the number of reflections, written by participants within the LLM (Human) intervention group, containing a theme. In total, using a mixed-methods approach (\ref{sec:thematic_analysis}), we annotated $102$ ($= 34*3$) reflections each from participants within the LLM and Human conditions (34 participants enrolled in each of the LLM and Human conditions, with each participant providing reflections for the 3 questions relevant to DV3). Note that a reflection could contain multiple themes.}
    \begin{tabular}{@{}p{0.17 \textwidth}|p{0.32\textwidth}|p{0.32\textwidth}|p{0.06\textwidth}|p{0.06\textwidth}@{}}
    \toprule
       
       \multicolumn{5}{@{}p{\textwidth}@{}}{\textbf{Quiz questions relevant to DV3}} \\

       \midrule
       
       \multicolumn{5}{@{}p{\textwidth}@{}}{Day 1: Based on the responses you read, do you think it is easier for someone who has been clean for 2 years to relapse compared to someone who has been clean for 10 years? Explain your answer.} \\
       \multicolumn{5}{@{}p{\textwidth}@{}}{Day 1: Is a full recovery possible?} \\
       \multicolumn{5}{@{}p{\textwidth}@{}}{Day 9: Is it possible to outgrow drugs? Explain your answer.} \\
       
        \midrule

    \textbf{Theme} & \textbf{Definition} & \textbf{Example} & \textbf{$n_L$} & \textbf{$n_H$} \\
    
    \midrule
        Fatality~\cite{MADDEN2019324,richard2020you,BAGLEY2023105} & Belief that full recovery from OUD is not possible; cravings never go away, and recovery involving another drug (i.e., MAT) is \textit{not} full recovery. & ``[...] it's not possible to fully recover, mentally, from the effects of it.'' -- (H1) & $35/102$ & $46/102$ \\ \hline
        
        Belief in Recovery~\cite{BAGLEY2023105} & Belief that full recovery is possible. Focuses on the improvement of the individual. & ``A full recovery is possible, but not without hard work and daily challenges to stay sober.'' -- (L4) & $80/102$ & $69/102$ \\ \hline
        
        Definition of Recovery & Focuses on explicitly defining or clarifying what `recovery' means -- defines it as individuals seeking and staying committed to treatment. & ``The [recovery] process is more about seeking treatment than outgrowing the addiction.'' -- (L4) & $41/102$ & $31/102$ \\ \hline
        
        Importance of a Strong Support System & Acknowledgment that a strong network of people, who provide direct, interpersonal support, is an essential component to full recovery. & ``It is near impossible to recover without a strong support circle.'' -- (H4) & $20/102$ & $12/109$ \\ \hline
        
        Moral Strength~\cite{chou2022stigma} & Belief that the onset of and recovery from OUD are associated with an individual's willpower. & ``[Recovery] depends on the person and how strong of a mindset they have.'' -- (H5) & $8/102$ & $9/102$ \\

    \bottomrule
    \end{tabular}
    \label{tab:codebook_DV3}
\end{table}
\begin{table*}
    \centering
    \caption{Few-shot with chain-of-thought prompt to automatically annotate themes or codes within participants' text-based reflections to the quiz questions.}
    \begin{tabular}{@{}p{\textwidth}@{}}
    \toprule 
    \emph{Few-shot with chain-of-thought prompt} \\
    You are an expert in text analysis and information related to opioid use disorder.
    First, please analyze the following examples where for a given question and answer pair, themes (or labels) that are best represented in the answer are identified using a pre-defined list. \\
    The pre-defined themes list, with the theme name and their corresponding definitions, is provided below. They are listed one on each line. \\\\

    \emph{Themes:} \\
    Theme 1: \textsc{[theme name]}; Definition: \textsc{[definition text]} \\
    Theme 2: \textsc{[theme name]}; Definition: \textsc{[definition text]} \\
    Theme 3: \textsc{[theme name]}; Definition: \textsc{[definition text]} \\
    Theme 4: \textsc{[theme name]}; Definition: \textsc{[definition text]} \\
    Theme 5: \textsc{[theme name]}; Definition: \textsc{[definition text]} \\\\
    
    \emph{Examples:} \\
    Following are examples, one on each line. Each example contains the \emph{question} and \emph{answer} pair, and the \emph{theme} (from the above list of themes) present within the \emph{answer}. We also provide a \emph{reasoning} for why the answer contains a particular theme. \\
    Question 1: \textsc{[question text]}; Answer 1: \textsc{[answer text]}; Theme: \textsc{[theme text]}; Reasoning: \textsc{[reasoning text]} \\
    Question 2: \textsc{[question text]}; Answer 2: \textsc{[answer text]}; Theme: \textsc{[theme text]}; Reasoning: \textsc{[reasoning text]} \\
    Question 3: \textsc{[question text]}; Answer 3: \textsc{[answer text]}; Theme: \textsc{[theme text]}; Reasoning: \textsc{[reasoning text]} \\
    Question 4: \textsc{[question text]}; Answer 4: \textsc{[answer text]}; Theme: \textsc{[theme text]}; Reasoning: \textsc{[reasoning text]} \\
    Question 5: \textsc{[question text]}; Answer 5: \textsc{[answer text]}; Theme: \textsc{[theme text]}; Reasoning: \textsc{[reasoning text]} \\\\

    Task: Given what you learned from the examples above, your task is to determine what themes are present within the following answer. An answer can contain multiple themes. Identify the most salient ones. Only exact theme names, from the above provided list, should be generated. New or modified themes, or incomplete themes should not be included. \\
    Generate a comma-separated list of theme names present within the answer.\\
    Also provide a reasoning for your output, quoting excerpts from the answer. \\
    Do not restate the answer. \emph{Please think through step by step.} \\\\

    Question: \textsc{[question text]}\\
    Answer: \textsc{[answer text]} \\
    Themes: \\
    Reasoning: \\
    \bottomrule
    \end{tabular}
    \label{tab:chain-of-thought-prompt}
\end{table*}
\begin{table}
    \centering
    \caption{\emph{Post-intervention change in Control.} Kruskal Wallis H-tests (or non-parametric one-way Anova tests) were performed to see if the pre- and post-intervention distributions of participants' attitudes toward the three DVs were significantly different. For both the single and longitudinal exposure setups, the differences between pre- and post-intervention attitudes were not significantly different ($p$-value ($p$) > $0.05$; the pre-registered significance threshold).}
    \begin{tabular}{@{}l|r|r@{}}
    \toprule
    \multicolumn{3}{l}{\emph{Single exposure setup}} \\
    \midrule
       \textbf{DV}  & \textbf{H-statistic} & \textbf{$p$} \\
    \hline
       DV1 & $3.487$ & $0.062$ \\
       DV1 & $8.635$ & $0.099$ \\
       DV3 & $0.873$ & $0.350$ \\
    \bottomrule 
    \multicolumn{3}{l}{\emph{Longitudinal exposure setup}} \\
    \midrule
        
       DV1 & $1.229$ & $0.541$ \\
       DV2 & $1.544$ & $0.462$ \\
       DV3 & $4.628$ & $0.099$ \\
    \bottomrule
    \end{tabular}
    \label{tab:control-anova}
\end{table}
\begin{table}
    \centering
    \caption{\emph{Impact of consuming information on OUD outside the intervention.} Post-hoc analysis performed to assess whether consuming information on OUD, outside of our provided interventions, had an impact on participants' change in attitudes. We introduced an additional fixed effect, i.e., \texttt{ext\_engagement} (which captured
    participants' Yes/No response to the question: Have you explicitly read about OUD online, in the news, or on the internet since you enrolled in this study?), to our original linear mixed-effects model (\ref{sec:lmm}). The \texttt{ext\_engagement} fixed effect did not have a statistically significant impact on post-intervention change in attitudes (i.e., $\delta Y$): $p$-value ($p$) > $0.05$; the pre-registered significance threshold.}
    \begin{tabular}{@{}l|r|r@{}}
    \toprule
    \multicolumn{3}{c}{\emph{Single exposure setup}} \\
    \midrule
       \textbf{DV}  & \textbf{t-statistic} & \textbf{$p$} \\
    \hline
       DV1 & $-0.951$ & $0.342$ \\
       DV2 & $-1.815$ & $0.070$ \\
       DV3 & $0.895$ & $0.371$ \\
    \bottomrule 
    \multicolumn{3}{c}{\emph{Longitudinal exposure setup}} \\
    \midrule
       DV1 & $0.746$ & $0.457$ \\
       DV2 & $0.253$ & $0.800$ \\
       DV3 & $-1.591$ & $0.115$ \\
    \bottomrule
    \end{tabular}
    \label{tab:ext_engagement}
\end{table}

\subsection{Sample Size} \label{sec:sm_sample_size}

For Study (a), i.e., the single exposure setup, we referred to the most recent publicly available statistics to help estimate the required number of participants. In 2022, the National Center for Health Statistics reported that $58.5$\% of adults in the United States seek healthcare-related information on online platforms~\cite{cdcProductsData}. Number of adults in the United States, in 2022, was reported to be $260,046,087$. Therefore, in line with our recruitment criteria, our target population size was $152,126,960$. For a target population of $152,126,960$, with a $4$\% error margin and $95$\% confidence interval, we were required to recruit $601$ participants per intervention condition. Therefore, in total, we needed $1,803$ ($601*3$) participants across the three LLM, Human, and Control interventions. Accounting for the standard attrition rate of $20$ to $30$\%, we recruited $2,400$ participants in total.

We used a power analysis to estimate the number of participants needed for Study (b), i.e., the longitudinal exposure setup. For a medium-sized effect (Cohen's $d$ = $0.57$) in change in attitudes toward substance use disorders and people with the condition~\cite{koyi2018change}, a power analysis with a significance level of $0.05$, powered at $80$\%, indicated that we required 40 participants in each condition. Therefore, in total, we needed $120$ ($40*3$) participants across the three LLM, Human, and Control interventions. We recruited $150$ participants to account for a $25$\% attrition rate, commonly observed in longitudinal experiments~\cite{gustavson2012attrition}.

\subsection{Intervention Dataset} \label{sec:sm_dataset}

\texttt{Reddit-QA}~\cite{reddit-qa} consists of $150,436$ original posts with an OUD-related query and the associated comments -- \textit{acting as a proxy for human-written responses}. Posts containing a query were identified using a BERT-based question classifier~\cite{turc2019wellread}, trained on a Quora question-answer dataset. Laud et al. used the Pushshift Reddit API to collect the \texttt{Reddit-QA} dataset from January 2018 to September 2021. \texttt{Reddit-QA} consists of data from 19 OUD-related subreddits, e.g., \texttt{r/Methadone}, \texttt{r/OpiatesRecovery}, and \texttt{r/Heroin}, including those explored by prior work and having a large membership. The posts contain queries on a variety of OUD-related topics, e.g., drug sales, drug use, addiction treatment, withdrawal, side effects, and lifestyle.

\emph{Data Filtering.} We processed and filtered the \texttt{Reddit-QA} dataset to select a subset of $112$ posts required for our experiment. First, given our experimental setup, we removed posts that received no comments. Prior work has shown that users' engagement on Reddit, via upvoting and downvoting, is reflective of their interest and conformity to community norms~\cite{PMID:29024875}. In addition, based on the platform's design, highly upvoted posts are more likely to be viewed by people~\cite{}. Therefore, we filtered out posts that had a less than average score (i.e, \#upvotes - \#downvotes), which was 3 for the \texttt{Reddit-QA} dataset. With the help of public health experts and well-validated topic models, Laud et al. labeled posts within the dataset for 10 major categories. Only 4 categories were relevant to our 3 dependent variables, i.e., attitudes toward MAT, people with OUD, and OUD, in our experiment -- drug use, withdrawal, addiction treatment, and lifestyle. Therefore, we only considered posts corresponding to these. This filtering process left us with $13,825$ posts. We then performed stratified random sampling, of these filtered posts, to select $112$ such that there was a balanced and diverse representation of topic categories and subreddits. During our Pilot (refer to~\ref{sec:sm_pilot}), participants echoed that, per intervention exposure, we should provide a mix of uplifting and distressing narratives. The first two authors read through the filtered $112$ posts to make sure that they contained a good balance of uplifting and distressing narratives. 

\textit{Human-generated Responses for Human Intervention.} For the filtered $112$ posts, described above, we then considered the top-most voted comment as the representative human-written response to the post. \texttt{Reddit-QA} belongs to a time period before the availability of democratized LLMs, e.g., ChatGPT (released in 2022), making it difficult for lay individuals to use them as writing assistants for posting content on Reddit -- therefore, ensuring that the comments within \texttt{Reddit-QA} are majorly human-written. The decision to consider the top-most voted comment was motivated by how Reddit drives user experiences, i.e., highly upvoted comments are ranked higher and are more likely to be viewed~\cite{mediumRedditRanking} -- it allowed us to mirror user interactions on Reddit and explore the impact of higher engagement, more likely to be seen human-written responses on people's attitudes.

\textit{LLM-generated Responses for LLM Intervention.} For the LLM intervention, we used \texttt{GPT-4} to get LLM-generated responses to the $112$ filtered \texttt{Reddit-QA} posts. \texttt{GPT-4} has outperformed several state-of-the-art systems on a variety of natural language generation tasks~\cite{openai2024gpt4}. Moreover, its ability to follow user intent has substantially improved over the prior LLMs. In particular, we used \texttt{gpt-4-0613}, which was the most stable and capable text generation model available at the time of this work. Following prior work, we chose a sampling temperature of $0.7$ -- it has been used to generate varied and creative responses, suitable to be posted on Reddit~\cite{10.1145/3544548.3581196}. 
We used prompt design to craft a textual input (Table~\ref{tab:gpt4-prompt}) guiding \texttt{GPT-4} to generate responses. We referred to well-adopted prompt engineering guidelines~\cite{prompt,promptingguideElementsPrompt} to draft this input. Broadly, it is recommended to include (a) simple task-specific instructions, (b) context, (c) question of interest, and (d) output format. Following these recommendations, we assigned \texttt{GPT-4} the persona of an active Reddit user who regularly seeks and provides information on OUD. We then prompted it to generate a response corresponding to a Reddit post containing an OUD-related query. The Reddit posts were fed as is from the earlier described and filtered \texttt{Reddit-QA} dataset. For additional context, we provided the post's subreddit name and description. We assigned the persona of a Reddit user because our goal was to investigate whether LLMs are fit to support the online health community ecosystem as information providers.

Table~\ref{tab:sample_dataset} lists paraphrased examples of posts, human- and LLM-generated responses within our intervention dataset. Raw descriptive statistics of the dataset are provided in Table~\ref{tab:data_statistics}.

\subsection{Pilot}\label{sec:sm_pilot}

We conducted a pilot, with three graduate students, to understand logistics of the intervention (i.e., the reading task). The three graduate students have expertise in social computing. Specifically, they have studied online communities and LLM-powered systems in the context of healthcare. To run this pilot, we randomly sampled $10$ posts from the filtered \texttt{Reddit-QA} dataset, along with their top-most voted comment. We then asked the three pilot participants to read these 10 posts (containing an OUD-related query) and the corresponding human-written response (i.e., the top-most voted comment). Following the completion of this reading task, the participants provided feedback on different aspects of the intervention such as task duration, content provided to read, attention checks, and accessibility. We then used their feedback to improve our study design. Following is a brief description of the feedback provided:

\textit{Task Duration:} All three participants expressed that the duration of the reading task or the intervention (i.e., engagement with either LLM- or human-written responses) should not be more than 15 minutes. They provided this feedback considering several factors: (1) type of content; based on the nature of this work, the interventions could contain distressing narratives (e.g., mentions of relapse or overdose), (2) reader attention span, and (3) potential exposure to misinformation. Our participants were able to read 8, out of the 10 query-response pairs provided, within 15 minutes. Therefore, in our final experiments, the intervention dosage was limited to 8 human- and 8 LLM-generated responses -- provided once in Study (a) and daily for 14 days in Study (b) -- for the Human and LLM intervention groups respectively. 

\textit{Interactive Platform:} In the pilot, the participants received the intervention using a Word document containing the 10 query-response pairs in continuity, one after the other. They found the length of the document to be overwhelming and requested for an interactive platform where they could engage with each query-response pair separately; similar to the design of Reddit. Based on this feedback, we designed an interactive interface (Figure~\ref{fig:interface}) to support participant engagement with the content. Refer to \ref{sec:sm_interface} for more details on the design/features of the interface.

\textit{Type of Content:} Given the nature of this work, the query-response pairs that the pilot participants read contained some distressing narratives on opioid use disorder, e.g., mentions of relapse, overdose, and struggles with social re-integration. To limit consumer distress, the pilot participants stressed on having a mix of uplifting and discouraging narratives during the intervention. In addition, the participants recommended to provide a glossary explaining the meaning of nuanced keywords or phrases (e.g., ``cold turkey'', ``12-step meeting'', and ``post-acute withdrawal syndrome''), which frequently appeared in the query-response pairs and may not be known to a broader audience.

\textit{Attention Checks:} To gauge reader attention and engagement, we asked generic True-False questions at the end of the reading task. For example, one of the questions we asked our pilot participants was: ``Is suboxone a medication-assisted treatment?'' After the pilot, the participants expressed that the attention-check questions should be (1) directed towards the query-response pairs read earlier and (2) complex/involved, i.e., subjective and difficult to search online. 

\subsection{Interface to Receive Treatment} \label{sec:sm_interface}

Based on the findings from our pilot (\ref{sec:sm_pilot}), we designed an interactive interface using which the participants  received the intervention. Following is a description of its key features:

\begin{enumerate}

\item \textit{Home Page:} On the very first page (Figure~\ref{fig:interface} (a)), we provided detailed descriptions about (a) the different components of the interface (e.g., instructions on how to use the glossary feature, the helpline panel, and buttons to read through the query-response pairs and access the quiz), (b) what not to do, and (c) participant compensation. We advised the participants to not use Generative AI-based search engines or conversational agents. We also constrained them from consuming any OUD-related information, outside of our provided intervention, during their participation. Additionally, we reiterated that the participants can withdraw their consent at any time, even after completing their participation, with no negative consequences.

\item \textit{Helpline Panel:} Due to the presence of distressing narratives within the query-response pairs, we added a helpline panel (Figure~\ref{fig:interface} (b)) that was visible and accessible throughout participant engagement. The panel listed an online resource to find nearby treatment facilities and the National Helpline as recommended by the Substance Abuse and Mental Health Services Administration (SAMHSA). We also provided contact information of the first author in case the participants had any questions or concerns about the study.

\item \textit{Glossary:} Based on our pilot, we decided to provide short descriptions for nuanced keywords or phrases, which may not be known to a broader audience -- e.g., ``suboxone'', ``cold turkey'', and ``post-acute withdrawal syndrome''. These domain-specific keywords were formatted in \textit{italics}, wherever they appeared in text, and the participants could hover over or tap them to access their description (Figure~\ref{fig:interface} (d)). We referred to descriptions provided by clinically-verified sources, e.g., SAMHSA, Centers for Disease Control and Prevention (CDC), and National Institute on Drug Abuse (NIDA), to curate the glossary.

\item \textit{Quiz:} After completing the reading task, we prompted the participants to take a mandatory quiz (Figure~\ref{fig:interface} (g)), which helped us gauge their attention to the intervention. Informed by our pilot, we included subjective, reading comprehension style questions. Instead of asking broad, generalized or searchable questions, we consciously decided to ask questions that prompted the participants to recall or comment on specific details they had read earlier within the responses. For example, one of the questions was: ``\textit{Based on the responses you read, do you think it is easier for someone who has been clean for 2 years to relapse as compared to someone who has been clean for 10 years? Explain your answer.}'' We also framed some of these questions as a proxy for participant attitudes toward the 3 DVs, i.e., attitudes toward MAT, attitudes toward people with OUD, and attitudes toward OUD.

\item \textit{Activity Tracker:} For additional attention checks, we tracked participant activity on the interface. We tracked (1) time taken to read each query-response pair, (2) number of times participants read through each query-response pair, (3) number of times participants used the glossary feature (i.e., hovered over the italicized terms), (4) number of times participants accessed the helpline panel, and (5) time taken to complete the quiz. This helped us identify and flag participants who did not read the content thoroughly, i.e., they either spent too little time completing the reading task or provided nonsensical answers to the quiz questions.

\item \textit{Disable Copy-Paste:} We disabled copy-paste functionality across the interface. This decision was made to make it difficult for participants to engage with Generative AI-powered systems or search engines to help answer the attention check quiz questions or to trace the source of the query-response pairs.

\end{enumerate}

\subsection{Dependent Variables} \label{sec:dv_survey}

\paragraph{DV1: Attitudes toward MAT.} Medication-assisted treatment (MAT), i.e., the use of Methadone or Suboxone -- the two most common forms of MAT for OUD in the United States, represents the gold standard of treatment approved by the Food and Drug Administration for OUD~\cite{samhsaMedicationsSubstance}. Despite its efficacy and widespread use, MAT continues to be stigmatized in the offline world. Not only does the general public hold negative attitudes toward MAT, institutions and organizations such as pharmacists and public health workers perceive patients on MAT programs as untrustworthy, incompetent, and `junkies'~\cite{doi:10.1177/1178221816685087}. Such perceptions stem from \emph{intervention-based} stigma, i.e., \emph{MAT is simply replacing one drug with another}. They act as barriers to harm reduction -- people report reluctance to initiate or continue treatment, distrust toward the healthcare system, and lowered self-esteem~\cite{doi:10.1177/1178221816685087}. 

Moreover, prior works in social computing have presented evidence of myths or misinformation associated with MAT in the online contexts~\cite{doi:10.1177/0022042620986508,doi:10.1080/10826084.2022.2091786,elsherief2021characterizing}. ElSherief et al. found pronounced presence of the myth, MAT is simply replacing one drug with another, and its linguistic variants on Twitter, Reddit, YouTube, and Drugs-Forum~\cite{elsherief2021characterizing}. Twitter is also home to multiple conspiracy theories related to MAT -- people describe them as ineffective and as a scam to generate money for big pharmaceuticals~\cite{doi:10.1177/0022042620986508}. People on MAT are further dismissed online and not recognized as those recovering from OUD~\cite{doi:10.1080/10826084.2022.2091786}. Recently, with democratized access to LLM-based chatbots (e.g., ChatGPT), people are using them as question-answer systems to seek healthcare related information (e.g., for diagnosis and treatment)~\cite{doi:10.2147/JMDH.S447368}. Mittal et al. found that \texttt{GPT-4} propagates the MAT-related myth, though not as prominently as online communities on Reddit~\cite{icwsm2025}. 

Such online misinformation on MAT increases barriers to treatment. As people often seek non-conventional ways to OUD recovery, i.e., via anonymized online communities or readily available LLM-based chatbots~\cite{info:doi/10.2196/15293,scientificamericanChatbotsCould,tustumi2023future,doi:10.2147/JMDH.S447368}, it is essential to understand how these everyday online engagements impact people's attitudes toward MAT. To this end, we adopt statements from the widely used Attitudes Toward Methadone Questionnaire~\cite{brown1975methadone} for our participants to self-report their opinions on MAT (i.e., DV1), before and after the interventions. The questionnaire has been used to understand attitudes of first responders, people on MAT recovery programs, and program administrators toward MAT~\cite{KRUIS2021108485,brown1975methadone}. The questionnaire consists of statements about MAT's efficacy (i.e., whether or not MAT simply replaces one drug with another) and potential for harm reduction (i.e., societal re-integration and reduction in crime). DV1 was measured by a combined response to 6 statements: (1) MAT drugs are safe drugs, (2) Once you are on MAT you have to keep taking it, (3) MAT allows people who have suffered from OUD in the past to lead a normal life, (4) MAT is the best way to quit opioids, (5) MAT is effective at reducing future crime, and (6) MAT is as addictive as opioids. Responses to these statements followed a 5-point Likert scale (ranging from strong agreement ($= 1$) to strong disagreement ($= 5$)). Statements (1), (3), (4), and (5) were reverse coded. Participant responses were averaged across the 6 statements. Higher scores were indicative of less stigmatized attitudes toward MAT. 

\paragraph{DV2: Attitudes toward people with OUD.} Public, both offline and online, associates derogatory terms (e.g., `addicts' or `junkies') when describing people with OUD~\cite{doi:10.1080/10826084.2022.2091786,ghosh2022does}. Over the years, scholars in public health have summarized three measures -- \emph{dangerousness}, \emph{blame}, and \emph{social distance} -- to assess stigmatized perceptions toward people with OUD~\cite{KRUIS2021108485,yang2019new,doi:10.1080/10826084.2022.2091786}. Dangerousness labels people with OUD as a threat to safety of the community. Blame encompasses the belief that people with OUD are responsible for the onset or exacerbation of their own drug using habits and any negative consequences associated with it. Social distance refers to a preference to avoid people with OUD in social settings. People with OUD often turn to anonymized online communities on Reddit, as safe spaces, to avoid the public shame in offline contexts~\cite{Balsamo_Bajardi_De_Francisci_Morales_Monti_Schifanella_2023,doi:10.1080/10826084.2022.2091786}. However, prior works have shown that posts/comments on Reddit promote the stigmatized measures of dangerousness, blame, and social distance when describing people with OUD. For example, Mittal et al. found that Reddit promoted the measure of blame with linguistic variants of ``People with OUD are suffering from a self-imposed condition maintained through the lack of moral fiber''~\cite{icwsm2025}. Such attitudes, when endorsed publicly on online forums, can negatively influence people's behavior. Moreover, this can have negative consequences for people with OUD, a group that already experiences widespread social disadvantages and systemic inequities. Consequently, facing potential or real discrimination in employment or education, coupled with the fear of being labeled as `dangerous', `weak', or `addicts', people with OUD may avoid the online ecosystem to self-disclose or seek help and treatment~\cite{wahl_news_2003} -- isolating them further. In addition, with the increase in adoption of LLMs for health information seeking~\cite{scientificamericanChatbotsCould}, it is imperative to understand how people with OUD are represented within these alternate online systems. 

To capture participants' attitudes toward people with OUD (i.e., DV2), we used opinion statements from the Brief Opioid Stigma Scale and other surveys on social stigma~\cite{yang2019new,KRUIS2021108485}. Statements captured the 3 measures of stigmatized perceptions, dangerousness, blame, and social distance, toward people with OUD. DV2 was measured by a combined response to 9 statements: (1) Persons with opioid use disorder are usually responsible for their own condition (\emph{blame}), (2) The job performance of people with opioid use disorder is largely impacted (\emph{dangerousness}), (3) Those addicted to opioids are self-harming persons who exacerbate their own condition (\emph{blame}), (4) People who use opioids are a threat to the safety of our community (\emph{dangerousness}), (5) It would be difficult for me to develop a friendship with someone who uses opioids (\emph{social distance}), (6) People with opioid use disorder are more susceptible to other mental health issues (\emph{social distance}), (7) Most people would willingly accept someone who has been treated for opioid use disorder as a close friend (\emph{social distance}), (8) It would bother me to live near a person who used opioids (\emph{social distance}), and (9) People's reflexes and coordination are not good when they are taking opioids (\emph{dangerousness}). Participant responses followed a 5-point Likert scale (ranging from strong agreement ($= 1$) to strong disagreement ($= 5$)). Statement (7) was reverse coded and participant responses were averaged across the 9 statements. Higher scores were indicative of less stigmatized attitudes toward people with OUD. 

\paragraph{DV3: Attitudes toward OUD.} Opioid addiction is falsely recognized as a \emph{willful choice}, and not a disease treatable with medicine~\cite{doi:10.1080/10826084.2022.2079133}. This stigmatized perception leads to \emph{condition-based} stigma~\cite{MADDEN2019324}, where people, including the general public, healthcare providers, and professionals, do not believe in the potential for full recovery. It embraces the opinion that opioid using persons are confined to a life of drug use -- that they will never get `clean'. In addition, since OUD is perceived as having a moral etiology~\cite{richard2020you}, the onset of and recovery from the condition are linked to an individual's willpower or moral strength. In this view of OUD, complete abstinence from drugs is seen as the most common and effective route to recovery. It further sparks \emph{intervention-based} stigma in which treatment programs are seen as a `crutch' (or a weak moral choice) and people recovering through them are frowned upon -- people on MAT are not recognized as those recovering from opioid use~\cite{doi:10.1080/10826084.2022.2091786}. ElSherief et al.~\cite{info:doi/10.2196/44726} found support for the moral etiology framework and complete abstinence on online forums, namely, Reddit, Twitter, YouTube, and Drugs-Forum. The online public presence of condition-based stigma and support for moral etiology can negatively shape public perception, harm individuals with OUD, and heighten barriers to treatment.

To examine participants' attitudes toward OUD (i.e., DV3), we used opinion statements from surveys on assessing condition-based stigma~\cite{KRUIS2021108485}. DV3 was measured by a combined response to 6 statements: (1) Full recovery from opioid addiction is not possible, (2) Moral strength plays a large part in the cause of opioid use disorder, (3) A reason why opioids have caused problems is that people can get them too easily, (4) Most people who become addicted to opioids are addicts for life, (5) Treating persons addicted to opioids seems futile, and (6) It's harder to concentrate when you're taking opioids. Participant responses to these statements followed a 5-point Likert scale (ranging from strong agreement ($= 1$) to strong disagreement ($= 5$)). Responses were averaged across the 6 statements. Higher scores were indicative of less stigmatizing attitudes toward OUD.

\subsection{Prolific Study Logistics}
Before launching the two studies, we ran a pilot of Study (b), i.e., the longitudinal exposure setup on Prolific. We recruited 6 participants from Prolific (3 per LLM and Human interventions). This helped us make Prolific-specific logistical decisions around participant retention, dropout criteria, and compensation. We also referred to resources provided by Prolific to successfully run longitudinal experiments~\cite{prolificLongitudinalMultipart}. Accordingly, we followed the study logistics described below while running the single and longitudinal exposure setups on Prolific:

\textit{Time:} We invited participants to complete the different phases of the two studies, e.g., the pre-intervention survey, daily interventions, mid- and post-intervention surveys at 11:00 AM Eastern Time. This allowed us to launch the different phases during office hours, when Prolific participants are the most active~\cite{prolificWhenProlific}, across the different time zones within the United States. 

\textit{Participation Reminder:} Since both the single and longitudinal exposure setups required participants to complete multiple phases (e.g., the pre-intervention survey, daily interventions, mid- and post-intervention surveys), spread across days, we sent friendly reminders twice -- at 6:00 PM Eastern Time the same day and at 9:00 AM Eastern Time the next day -- to continue participation. This helped us retain participants. In the post-intervention survey, all of our participants reported that the reminders were helpful -- 
``\emph{They [reminders] helped me stay on track. I enjoyed the notices and couldn't wait to login and continue my participation (participant enrolled in the longitudinal exposure setup; LLM intervention)}.''
We sent these reminders using the internal messaging system provided by Prolific.

\textit{Participant Contact:} We contacted participants, through Prolific's internal messaging system, to (a) send reminders (described above) and (b) request for a re-take (Table~\ref{tab:prolific-message}). Participants who failed attention checks, i.e., took too little time to go through the daily interventions (much less than the average) or provided gibberish answers to the post intervention daily quiz questions, were requested to take the intervention again.

\subsection{Participant Attrition}\label{sec:attrition}
Participants were free to discontinue participation at any point during the two studies. To ensure data quality, we ourselves removed participants who failed attention checks, twice, with multiple messages sent to re-take. We informed them, via Prolific, before removing them from our participant pool. Participants were compensated for all the phases of the studies they completed. 
For the single exposure setup, $2,141$ ($772$ in Control, $696$ in Human, and $673$ in LLM) out of the $2,400$ recruited participants completed the experiment, resulting in an attrition rate of $10.79$\%.  
For the longitudinal exposure setup, $107$ ($39$ in Control, $34$ in Human, and $34$ in LLM) out of the $150$ recruited participants completed the experiment. This resulted in an attrition rate of $28.67$\%. Out of the $43$ participants who dropped out, 20 ($13.33$\%) were unresponsive and the remaining 23 ($15.33$\%) were removed by us as they repeatedly failed attention checks.  

\subsection{Participant Demographics} \label{sec:demographics}
Here we describe the participant demographics for Study (a) and Study (b), excluding those who dropped out in between.

\paragraph{Study (a).} In the single exposure setup, $57.07$\% participants identified as female, $40.88$\% identified as male, $1.82$\% identified as non-binary, and the remaining $0.23$\% preferred not to report their gender. All the participants were from the United States, which was one of our recruitment criteria. $16.02$\% participants belonged to the 18-25 years of age bracket, $15.55$\% participants belonged to the 26-30 age bracket, $31.90$\% participants belonged to the 31-40 age bracket, $24.29$\% participants belonged to the 41-50 age bracket, and $12.24$\% participants belonged to the 51-60 age bracket. $42.46$\% participants identified as democrats, $29.99$\% identified as republicans, and the remaining $27.55$\% identified as somewhere in between. $87.53$\% of the participants reported having some familiarity with OUD or MAT. $85.71$\% of the participants self-reported that they use online platforms, such as online communities on Reddit, Facebook and search engines, or Generative AI-based conversational agents (e.g., ChatGPT) to seek healthcare-related information. 

\paragraph{Study (b).} In the longitudinal exposure setup, $62.61$\% participants identified as female, $35.52$\% identified as male, and the remaining $1.87$\% identified as non binary. Again, based on our recruitment criteria, all the participants were from the United States. $12.15$\% participants belonged to the 18-25 age bracket, $27.10$\% participants belonged to the 26-30 age bracket, $33.64$\% participants belonged to the 31-40 age bracket, $19.63$\% participants belonged to the 41-50 age bracket, and the remaining $7.48$\% participants belonged to the 51-60 age bracket. $45.79$\% participants identified as democrats, $23.36$\% participants identified as republicans, and the remaining $30.85$\% participants identified as somewhere in between. $73.83$\% participants self-reported having some familiarity with OUD or MAT. $84.11$\% participants reported that they use online platforms or Generative AI-based conversational agents to seek healthcare-related information. 

\subsection{Linear mixed-effects model}\label{sec:lmm}

To test the main effect of the interventions on the 3 dependent variables, for both single and longitudinal exposure setups, we performed a between-condition analysis using a linear mixed-effects model. For each DV, we estimated the change in attitudes from pre- to post-intervention, i.e., $\delta Y = (Y_{post} - Y_{pre})$ where $Y_{post}$ and $Y_{pre}$ represent the aggregated score for each DV post- and pre-intervention respectively. We leveraged the \texttt{lmer} function within the \texttt{lmerTest} library, provided in $R$, for this analysis. The Control condition was used as a reference. In addition to the \texttt{condition} variable (i.e., LLM, Human, or Control), we included additional pre-registered fixed-effects, namely, demographics such as political leaning (\texttt{pol\_leaning}) and education level (\texttt{edu\_level}), participants' prior association with people with OUD (\texttt{oud\_individual\_association}), and participants' reliability on online information (\texttt{online\_info\_reliability}). Also, given that our intervention assignment was based on stratified randomization and blocks, we introduced \texttt{block} as a random effect, to account for any variance across the blocking factors, i.e., age, gender, familiarity with OUD, reliance on online platforms for health information seeking, and aggregated baseline attitudes toward the 3 DVs. 
\begin{align*}
    \delta Y &\sim \texttt{condition} + \texttt{pol\_leaning} + \texttt{edu\_level} + \texttt{oud\_individual\_association} + \\ &\texttt{online\_info\_reliability} + (1|\texttt{block})
\end{align*}

\subsection{Thematic Analysis of Post-intervention Text-based Reflections}\label{sec:thematic_analysis}

As previously described (\ref{sec:sm_interface}), participants were required to complete a mandatory quiz containing subjective, reading comprehension-based questions after finishing the reading task or the daily intervention. Beyond serving as an attention check, some of the questions were intentionally designed to act as a proxy for assessing participants' attitudes toward the three dependent variables, through their open-ended, text-based reflections on them. 

\textbf{Codebook.} For the longitudinal exposure setup, we conducted a thematic analysis~\cite{roberts2019attempting} of these text-based reflections to examine  participants' perceptions over the course of the exposure. To do so, we first identified quiz questions relevant to each DV. Additionally, we ensured that the selected questions were distributed across the 14 day duration of the longitudinal exposure setup. For DV1, we selected 3 relevant questions asked on days 2, 7 and 14 of the exposure. For DV2, we selected 4 questions asked on days 4, 8 and 14. For DV3, we selected 3 questions asked on days 1 and 9. For each DV, we prepared a separate codebook to annotate themes within participants' reflections to the selected questions. We began with a deductive coding~\cite{azungah2018qualitative} based approach by including themes relevant to the DVs and well-established in prior work. For example, for DV1 we included themes such as `Dependence'~\cite{doi:10.1177/0022042620986508,doi:10.1080/10826084.2022.2091786,elsherief2021characterizing,MADDEN2019324} to cover negative attitudes toward MAT. Following this, we randomly selected 5 participants from each of the Human and LLM conditions to hand-annotate their daily reflections and iteratively refine the codebooks. During the annotation process, we also introduced new, context-specific themes via inductive coding~\cite{thomas2003general}. Two authors, with prior experience in social computing and data science, led this effort. In total, they coded $100$ ($= (3 + 4 + 3)*10$) reflections. Per inter-rater agreement, the two annotators scored a Cohen's Kappa of $0.87$. Next, they held an in-depth discussion to resolve conflicts, refine definitions for themes with low agreement, and eliminate themes that were not prominent in the dataset. The final codebooks are summarized in Tables~\ref{tab:codebook_DV1},~\ref{tab:codebook_DV2}, and~\ref{tab:codebook_DV3}.

\textbf{Automated Theme Annotation.} We used the above described hand-annotated dataset as ground truth to automatically identify themes in the reflections provided by the remaining 58 participants (29 from each of the Human and LLM interventions). To do so, we made use of a few-shot learning~\cite{NEURIPS2020_1457c0d6} approach, with chain-of-thought prompting~\cite{Wei2022ChainOT}, to identify themes within a given text. Few-shot prompts that condition large language models on a few task-specific examples are shown to outperform supervised training paradigms~\cite{zhang2023interpretable}. In addition, a step-by-step reasoning, using a chain-of-thought prompt, describing how a sample belongs to a class facilitates classification~\cite{madaan-etal-2023-makes}. We modeled our problem as a multi-label classification task, as the reflections could include multiple relevant themes. A separate classification prompt, and therefore, a classifier was created to annotate themes within reflections for quiz questions relevant to each DV (hence, 3 prompts in total). Based on prior recommendations~\cite{Parnami2022LearningFF}, we provided 3 to 5 examples per theme, from the hand-annotated ground truth dataset, as few-shots for training within the prompts. We referred to prompt-engineering guidelines to create these classifiers~\cite{prompt,promptingguideChainofThoughtPrompting}. The prompt is provided in Table~\ref{tab:chain-of-thought-prompt}. We used the open-weights \texttt{Llama-3.1-70B}~\cite{metaIntroducingLlama} model to create the few-shot learners. We did not use a \texttt{GPT}-based model for classification because we leveraged \texttt{GPT-4} for the LLM intervention response generation (\ref{sec:sm_dataset}), and wanted to minimize overlap between the models used for classification and generation. Per previous work, we used a sampling temperature of $0$ as the task at hand -- classification -- requires more consistent, well-defined, and coherent outputs~\cite{temperature}.

\textbf{Classifier Validation.} First, we validated the classifiers for each DV using the annotated reflections from the ground truth dataset that were not included as few-shot examples. This validation dataset consisted of 19, 30, and 21 hand-annotated reflections for DV1, DV2, and DV3, respectively. We then compared the labels generated by the classifiers with our ground truth annotations. This resulted in a precision of $0.69$, a recall of $0.89$, and an F1-score of $0.78$ for DV1; a precision of $0.66$, a recall of $0.97$, and an F1-score of $0.79$ for DV2; and a precision of $0.74$, a recall of $1.00$, and an F1-score of $0.85$ for DV3. Next, we used the classifiers to get themes within the reflections provided by the remaining 58 participants (29 from each of the Human and LLM conditions), constituting our test dataset. We then conducted a second round of human validation, using a subset of reflections within the test dataset. For each DV, we considered reflections written by 14 randomly selected participants -- 7 each from the Human and LLM conditions -- forming our validation dataset. Therefore, the validation dataset consisted of 42 (14*3), 56 (14*4), and 42 (14*3) reflections for DV1, DV2, and DV3, respectively. One author hand-annotated these using the codebooks. The classifiers obtained the following performance metrics on comparing their generated labels against the manual annotations -- a precision of $0.74$, a recall of $0.89$, and an F1-score of $0.81$ for DV1; a precision of $0.73$, a recall of $0.93$, and an F1-score of $0.82$ for DV2; and a precision of $0.69$, a recall of $0.85$, and an F1-score of $0.76$ for DV3.

\subsection{Impact of Consuming Information on OUD Beyond Intervention}\label{sec:impact_ext_engagement}

After participating in the single and longitudinal exposure setups, we asked the participants whether they had consumed any information on OUD outside of our provided interventions. We asked the following Yes/No question in the post-intervention survey: \textit{Have you explicitly read about OUD online, in the news, or on the internet since you enrolled in this study?} As a sanity check, we performed a post-hoc analysis to investigate whether this external engagement had any impact on participants' change in attitudes. To do so, we introduced an additional fixed effect (\texttt{ext\_engagement}; which encoded participants' response to the earlier mentioned question) to our original linear mixed-effects model (\ref{sec:lmm}). 
\begin{align*}
    \delta Y &\sim \texttt{condition} + \texttt{pol\_leaning} + \texttt{edu\_level} + \texttt{oud\_individual\_association} + \\ &\texttt{online\_info\_reliability} + \texttt{ext\_engagement} + (1|\texttt{block})
\end{align*}

\noindent Next, we considered the impact of this additional \texttt{ext\_engagement} fixed effect on change in attitudes (i.e., $\delta Y$). The results of this analysis are summarized in Table~\ref{tab:ext_engagement}. Consuming OUD information, beyond our provided interventions, did not have a statistically significant impact on change in attitudes (i.e., $\delta Y$). This result was consistent across all three DVs and for both the single and longitudinal exposure setups. 

\end{document}